\pdfoutput=1

\documentclass[11pt]{article}

\usepackage[final]{acl}

\usepackage{times}
\usepackage{latexsym}
\usepackage{appendix}

\usepackage{amsmath,amsfonts,bm}
\usepackage{mathrsfs}

\def\eqref#1{equation~\ref{#1}}

\def\1{\bm{1}}

\DeclareMathAlphabet{\mathsfit}{\encodingdefault}{\sfdefault}{m}{sl}
\SetMathAlphabet{\mathsfit}{bold}{\encodingdefault}{\sfdefault}{bx}{n}

\def\gV{{\mathcal{V}}}

\def\sA{{\mathbb{A}}}

\def\emS{{S}}

\newtheorem{theorem}{Theorem}
\newtheorem{remark}{Remark}
\newtheorem{lemma}{Lemma}

\usepackage{algorithm}
\usepackage[noend]{algpseudocode}
\usepackage[pdftex]{graphicx}
\usepackage{enumitem}

\usepackage[T1]{fontenc}

\usepackage[utf8]{inputenc}

\usepackage{microtype}

\usepackage{inconsolata}

\usepackage{graphicx}
\usepackage{booktabs}
\usepackage{tabularx}
\usepackage{array}
\usepackage{pifont}
\usepackage{diagbox}

\allowdisplaybreaks
\title{OD-Stega: LLM-Based Relatively Secure Steganography via Optimized Distributions}

\author{
 \textbf{Yu-Shin Huang\textsuperscript{1}},
 \textbf{Peter Just\textsuperscript{1}},
 \textbf{Hanyun Yin\textsuperscript{2}},
 \textbf{Krishna Narayanan\textsuperscript{1}},
 \textbf{Ruihong Huang\textsuperscript{2}},
 \textbf{Chao Tian\textsuperscript{1}},
\\
 \textsuperscript{1}Department of Electrical and Computer Engineering, Texas A\&M University,\\
 \textsuperscript{2}Department of Computer Scicence and Engineering, Texas A\&M University
\\{
   {\ttfamily\{yushinh, peter601, hanyun\_yin, krn, huangrh, chao.tian\}@tamu.edu}
 }
}

\begin{document}
\maketitle
\begin{abstract}
We consider coverless steganography where a Large Language Model (LLM) is used to generate stego-texts in combination with arithmetic coding. An efficient method should embed secret bits in as few language tokens as possible while keeping the stego-text as natural as possible. We show that this problem is equivalent to maximizing the entropy of a replacement probability distribution of the next token generation, subject to a constraint on the divergence between the new distribution and the original one produced by the LLM. A closed-form solution is provided under either the KL divergence or the total variation constraint. Several important practical issues are also tackled: 1) An often-overlooked tokenization mismatch issue is resolved with a simple prompt selection approach, 2) The combination of the optimized distribution and the vocabulary truncation technique is considered, and 3) The incorporation of the proposed approach with existing (potentially non arithmetic coding based) techniques, e.g., the Discop technique.
\end{abstract}

\section{Introduction}
In a steganography system, Alice, the sender, aims to convey a secret message to Bob, the receiver. The carrier signal can take the form of text, image, audio, or video \citep{anderson1998limits, cox2007digital, provos2003hide}. In this work, we focus on natural language text messages as the carrier signals, and the resultant signal with the secret message embedded is therefore referred to as the stego-text. 
Alice transmits the stego-text to Bob via a channel monitored by an eavesdropper Eve. Eve wishes to determine whether there is a hidden message. Alice must ensure that the stego-text can be decoded correctly by Bob, and at the same time, guarantee with a high probability that Eve cannot detect the message.

Conventionally, steganography relies on an existing cover signal (cover text), and achieves steganography by making subtle changes imperceptible to Eve on the cover text 
\citep{topkara2006hiding, chang2010linguistic}. 
As LLMs have grown more powerful, coverless steganography has achieved significant gains in both capacity and stealth. By generating fluent, human-like text, LLM-based schemes can produce stego-text that is difficult to distinguish from natural language while embedding more secret information in shorter outputs than traditional cover-text-based methods \citep{fang2017generating, yang2018rnn, ziegler2019neural,xiang2017novel,dai2019towards,zhang2021provably,shen2020near,kaptchuk2021meteor,ding2023discop,de2024perfectly}.

Though not always the case (e.g., \cite{ding2023discop}), the underlying driver for LLM-based steganography is usually the arithmetic coding (AC) algorithm \citep{witten1987arithmetic}, which is an efficient data compression algorithm based on the idea that any finite-length finite-alphabet data sequence (e.g., text) can be mapped to a small interval in the range of $[0,1)$. In LLM-based steganography, Alice utilizes the AC \textbf{decoder}, together with the probability distribution produced by the LLM, to map the secret binary sequence to a stego-text. Bob can then recover the secret message by performing the AC encoding. Intuitively, the AC decoder performs sampling with the probability distribution given by the LLM, using the secret message bits as the driving randomness, where we assume the secret message has been pre-encrypted with a secret key shared between Alice and Bob but not Eve (see \cite{shen2020near,kaptchuk2021meteor}), and the encrypted message is an i.i.d. binary sequence.

In many scenarios, steganography security can be relaxed when Eve is computation-bounded (e.g., mobile devices), delay-constrained (e.g., streaming or time-sensitive tasks), or limited by societal constraints (e.g., censorship under constitutional protection). In such cases, Eve can be modeled as a weak detector, and correspondingly, the steganography security requirement can be relaxed. This consideration was in fact already implicit in several previous works invoking ``near-imperceptibility"  \citep{dai2019towards,shen2020near}. 
Generalizing this idea, we can replace the conditional probability distribution while ensuring deterministic, causal synchronization between Alice and Bob under relaxed security constraints. We refer to this approach
as \emph{relatively-secure steganography}. Since practical stego-texts always have \textit{finite lengths}, even perfectly secure approaches, e.g., \cite{zhang2024provably,ding2023discop,de2024perfectly}, will induce a non-zero probability of steganography being detected, and the relatively-secure steganography only needs to keep this probability under control. 

The generalized view suggests a fundamental tradeoff between the amount of secret bits one can hide in the stego-text and the detectability of steganography; the former consideration is usually measured by the embedding capability or embedding utilization in the literature \citep{dai2019towards,shen2020near, kaptchuk2021meteor,ding2023discop}. Improving the utilization is particularly important for LLM-based steganography, since the generative process in LLMs can become almost deterministic and therefore difficult to hide secrets. 

We formalize the entropy maximization problem under two different probability divergence measures, and provide closed-form solutions. We refer to this approach of choosing an optimized distribution as OD-Stega. OD-Stega provides an additional {\em design freedom} beyond the conventional perfectly secure steganography. By choosing the KL divergence (or total variation distance) constraint as a hyperparameter in the specific engineering application, it can either fully recover the underlying perfectly secure algorithm or take advantage of Eve's weakness when such knowledge is available.

In addition to the principled formulation outlined above, our work also tackles several practical issues. 
First, most previous LLM-based steganography relied inherently on a bijective tokenization assumption, which does not hold in practice. We provide a simple solution via LLM prompting selection.
Secondly, we combine OD-Stega with the existing technique of vocabulary truncation to reduce the computation complexity, and analyze the overall KL divergence of this strategy. Lastly, the proposed approach is universal and can be integrated into existing methods, and we specifically provide results on incorporating the proposed approach with Discop \cite{ding2023discop}.

\section{Preliminary} 

\subsection{LLM-based Steganography}

An LLM can provide an estimate for the conditional probability distribution for the next token, given the sequence of tokens preceding it 
\citep{vaswani2017attention,brown2020language,touvron2023llama}. To generate a natural language sequence, one can sample the tokens from these distributions in an autoregressive manner.

The secret message bit sequence $S$ in steganography is pre-encrypted with a secret key shared with Bob but hidden from Eve. Before encoding, Alice selects an initial prompt text $T_p$, independent of $S$, which determines the nature or semantics of the resulting stego-text.
To encode $\emS$, Alice uses an encoding function $f(T_p,\emS)$ to produce a sequence of tokens $\underline{x}_{i>0}=(x_1,x_2,x_3,\ldots)$, which is then converted to the corresponding stego-text ${T}_s$ via detokenizing. The prompt and the stego-text $(T_p,T_s)$ are sent on the public channel. Bob first converts ${T}_s$ into the token form $\underline{x}_{i>0}$, then uses a decoding function $g(\cdot)$ such that $g(T_p,\underline{x}_{i>0}) = \emS$.

In LLM-based steganography, both $f$ and $g$ rely on the same LLM. At time \( i \), an LLM takes the tokenized input \( \underline{x}_{i-1} = ( x_{-n_{p}-1}, x_{-n_{p}-2}, \cdots, x_{i-1}) \) as the prompt, where $\underline{x}_{0} = ( x_{-n_{p}-1}, x_{-n_{p}-2}, \cdots, x_{0})$ represents the tokenized sequence of $T_p$ and $n_p$ is the number of tokens in $T_p$. This produces the probability distribution \( P_{LLM} \) for the next token \( x_{i} \). We shall write it as \( P^{i}=P_{LLM}(\mathbf{Y} = x_{i} \mid \underline{x}_{i-1}) \), which is the conditional probability for the next token, given the proceeding tokens (in the context window). 

\subsection{Arithmetic Coding} \label{sec:AC}

\begin{figure}[t]
    \flushleft
    \includegraphics[width=1.02\columnwidth]{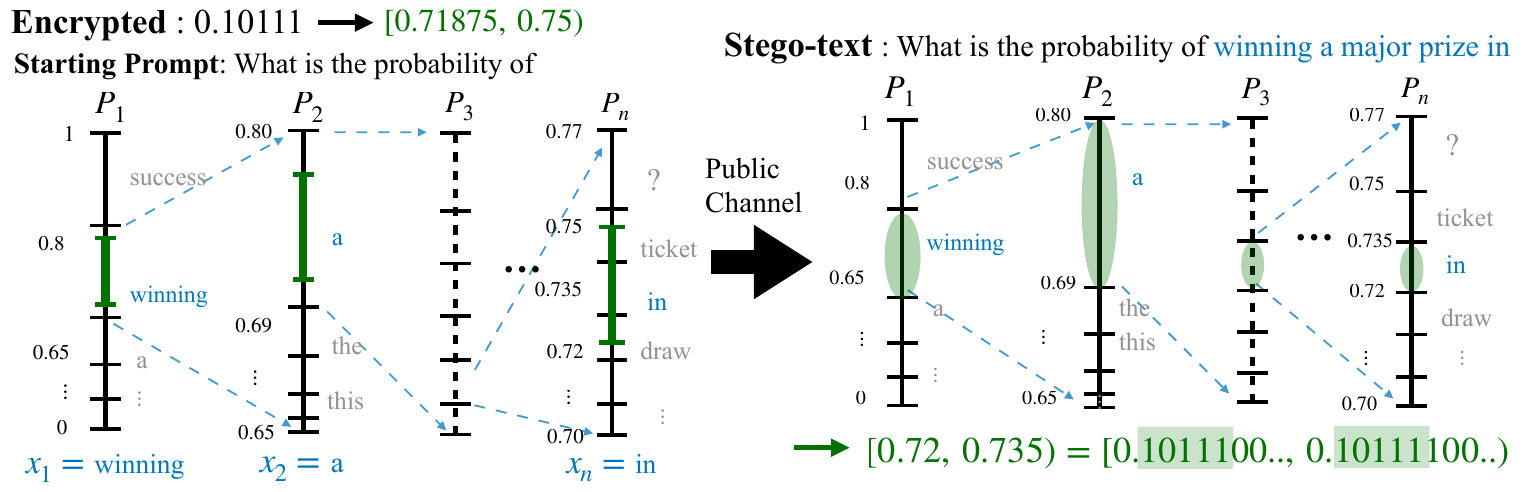}
    \caption{\small Example of AC in steganography: The sequence $10111$ can be represented as the interval $ \mathbf{I} = [0.101110000 \cdots_{2}, 0.1011111111\cdots_{2}) \simeq [0.71875, 0.75)$. We identify the range where this interval falls in the probability distribution $P^{i}$. }
    \vspace{-0.5cm}
    \label{fig:AC}
\end{figure}

Several authors have shown that Arithmetic Coding (AC), can be used together with language models to perform steganography \citep{ziegler2019neural, shen2020near, ivasenko2021information}. 
Typically, AC compresses the character in the sequence sequentially into a sequence of bits, and the decompressor can convert the sequence back to text. For steganography, an AC decoder can be viewed as a sampler in the set of natural language paragraphs using the secret message as a random seed, and since the secret message is uniformly distributed on the message set, the sampled text would look like natural language. There may be a small mismatch in the sampling probability as noted in \cite{ding2023discop}, however for a reasonably long sequence, this discrepancy becomes negligible. An illustrative example is given in Figure \ref{fig:AC}.

During decoding, Bob recognizes the starting token of the stego-text from the received text, and then derives the identical distribution from the same LLM with the starting prompt text. With those stego-text he receives, Bob retrieves the probabilities $P^{i>0}$ and reconstructs the bit sequence, until every bit is recovered.

\section{Proposed Methodology}
A well-known fact in data compression is that the expected minimum number of bits to represent a symbol following a probability $P$ is the entropy \(H(P)\) \citep{cover1991elements}, and AC is an algorithm that can compress at a rate close to this rate. The same relation holds for LLM-based steganography using AC, in the sense that the expected number of secret message bits that can be embedded for a given token position-$i$ is the entropy of the conditional distribution \(H(P^{i})\). For example, if a token has a conditional distribution of $\{\frac{1}{4},\frac{1}{4},\frac{1}{4},\frac{1}{4}\}$ on four possible token values, then 2 bits of secret message can be embedded in the stego-text.  

If Eve is a weak detector, then we can take advantage of the opportunity to make the conditional distribution $P$ more amicable for embedding secret message bits, i.e., choose a different distribution $Q$ with a higher \(H(Q)\). If $Q$ is close to $P$, we expect the generated stego-text to be nearly imperceptible to Eve. We model the detector strength of Eve via a divergence constraint $\delta$. By tuning $\delta$, we can recover a perfectly secure scheme or explicitly exploit Eve's limitations, leading to the formulation below.

\subsection{Optimized Distribution under Constraint} 
We formulate the following optimization problem.
{\small
\begin{align}
    \max_{Q_{j}^i, \; \forall j \in [1:N_i]} \quad & H(Q^{i}) = \sum_{j=1}^{N_i} -Q_j^{i} \log Q_j^{i} \label{eq:HPX}\\
    \text{subj. to} \quad & D(Q^{i} || P^{i}) \leq \delta \label{eq:KLconstraint} \\ 
    & Q_j^{i} \geq 0 , \quad \forall j \in [1:N_i] \label{eq:postiveconstraint} \\
    & \sum_{j=1}^{N_i} Q_j^{i} = 1 \label{eq:sumeq1_constraint}\\
    & Q_j^{i} = 0, \forall j \in \sA_i = [N_i+1: N] \label{eq:px_equal_0}
\end{align}
}where $N = |\gV|$ is the total number of symbols in the vocabulary. The objective function $H(Q^{i})$ is the standard Shannon entropy. The divergence, denoted as $D(Q^{i} || P^{i})$, could be KL divergence or TV (see discussion in \cite{dai2019towards,shen2020near}), and $\sA_i$ is the index set of elements with zero probability in $P^i$ (i.e., $P_j^{i}=0$). Without loss of generality, we assume that the elements in the vocabulary are in descending order of the probabilities $P^{i}$, and the number of nonzero elements in $P^{i}$ is written as $N_i$ (i.e., $N_i = N - |\sA_i|$).

In the optimization problem above, we seek to replace the natural language distribution probability distribution $P_i$ given by the LLMs with a new distribution $Q_i$ towards a larger entropy value (a more uniform distribution). This would allow for embedding a greater number of secret bits. The new distribution needs to be close to that of the natural language, which is ensured by the constraint in (\ref{eq:KLconstraint}). This optimization problem is convex as long as the divergence function in (\ref{eq:KLconstraint}) is convex.

\subsection{Optimal Probability Adjustment} \label{sec:opt_KL_strategy}
The main theoretical contribution of the work is shown in Theorem \ref{thm:Pxvalue_KL} and Theorem \ref{thm:Qvalue_TV}.
\begin{theorem} \label{thm:Pxvalue_KL}
When the divergence in (\ref{eq:KLconstraint}) is the KL divergence, the optimal solution $Q^{i}$ to the problem (\ref{eq:HPX})-(\ref{eq:px_equal_0}) is obtained by performing temperature scaling on $P^{i}$ when $\delta \in [0,\frac{1}{N_i} \sum_{j =1}^{N_i} \log(\frac{1}{N_i P_j^{i}}) ] $ 
{\small
\begin{align}
    Q_j^{i} &= \begin{cases}
        \frac{{P_j^{i}}^{\frac{1}{T}}}{\sum_{j=1}^{N_i}{P_j^{i}}^{\frac{1}{T}}}, &  \quad  \forall j \notin \sA_i  \label{eq:thm_KL_sol}\\
        0 , &  \quad  \forall j \in \sA_i 
    \end{cases} 
\end{align}
}for some $T \geq 1$ s.t. $D_{KL}(Q^{i}||P^{i}) = \delta$.  
Otherwise, 
{\small
\begin{align}
    Q_j^{i} &= \begin{cases}
        \frac{1}{N_i}, &   \quad  \forall j \notin \sA_i \notag \\
         0, &  \quad \forall j \in \sA_i \notag
    \end{cases}.
\end{align}}
\end{theorem}

\begin{remark} \label{remark:temp_scaling}
If the LLM generated distribution $P^{i}$ is defined via a softmax over logits $Z^{i}$, then applying temperature scaling $T$ to $P^{i}$ is equivalent to scaling the logits $Z^{i}$ by $\frac{1}{T}$ (see Appendix \ref{sec:pf-of-remark1} for a proof). 
\end{remark}

In Theorem \ref{thm:Pxvalue_KL} and Remark \ref{remark:temp_scaling}, we showed that the optimal adjustment $Q^{i}$ can be expressed by Equation (\ref{eq:thm_KL_sol}), which is mathematically equivalent to adjusting the temperature parameter in the LLM architecture. This finding offers an interesting theoretical justification for the practice of temperature tuning in language generation (without the steganography consideration), i.e., increasing the temperature maximizes the next token entropy under the KL divergence constraint.

\begin{lemma} \label{lemma:ulargerthan0}
    For $Q^{i}$ chosen as in (\ref{eq:thm_KL_sol}) and any $\delta \in [0,\frac{1}{N_i} \sum_{j =1}^{N_i} \log(\frac{1}{N_i P_j^{i}}) ] $, there exists a $T \geq 1$, such that the solution given in Theorem \ref{thm:Pxvalue_KL} satisfies the constraint (\ref{eq:KLconstraint}) with equality 
    {\small
    \begin{align}
        D_{KL}(Q^{i} || P^{i}) = \sum_{j=1}^{N_i} Q_j^{i} \log \left( \frac{Q_j^{i}}{P_j^{i}}\right) = \delta. \notag
    \end{align}
    }Moreover, $D_{KL}(Q^{i} || P^{i})$ is a monotonically increasing function of $T \geq 1$. 
\end{lemma}
The proofs of Theorem \ref{thm:Pxvalue_KL}, Remark \ref{remark:temp_scaling}, and Lemma \ref{lemma:ulargerthan0} are given in Appendix \ref{sec:pf-of-remark1}, \ref{pf:KL_solution} and \ref{sec:pflemma_ulargerthan0}, built on a careful analysis of the KKT conditions.
The specified $\delta$ only places a meaningful constraint within the range given in Theorem \ref{thm:Pxvalue_KL}. Otherwise, the KL constraint is too loose, and the optimal solution $Q^{i}$ defaults to a uniform distribution. 
It remains to solve for the temperature value $T$ that satisfies the KL constraint with equality. Lemma \ref{lemma:ulargerthan0} states that the KL divergence grows with increasing $T$, which allows us to numerically find \( T \) through a straightforward bisection search.

\begin{figure}
    \centering
    \includegraphics[width=\linewidth]{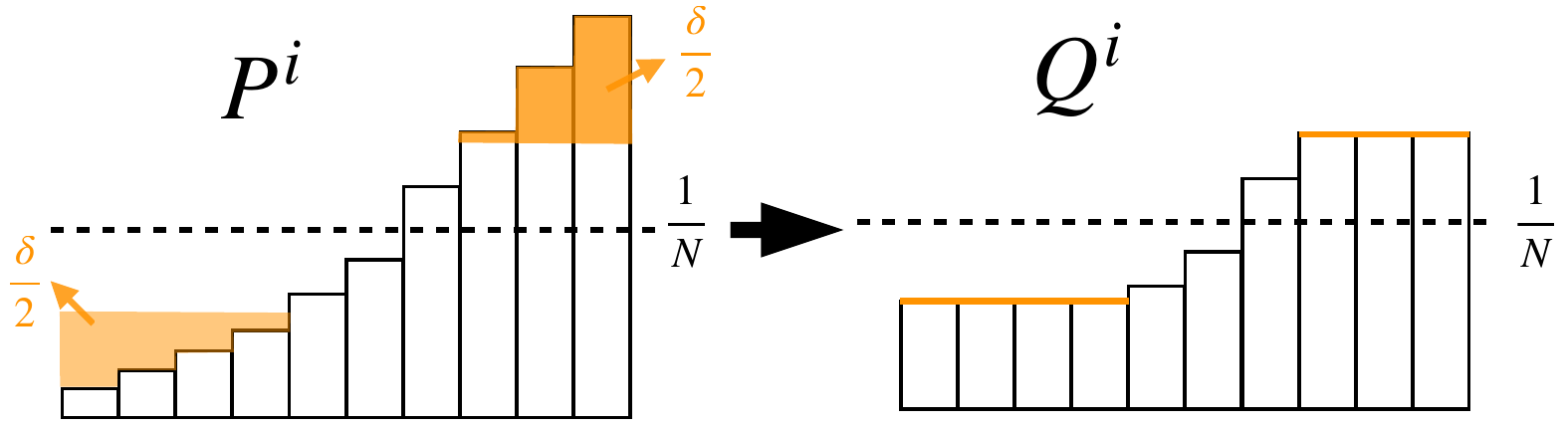}
    \caption{Optimal adjustment allocation under total variation constraint. }
    \label{fig:tv_illus}
\end{figure}

\begin{theorem}[informal] \label{thm:Qvalue_TV}
When the divergence in (\ref{eq:KLconstraint}) is the total variation distance (TV), the optimal solution $Q^{i}$ to the problem (\ref{eq:HPX})-(\ref{eq:px_equal_0}) is obtained by splitting the budget $\delta$ into half equally, half to increase the lower-end probabilities to a certain common value, and the other half to decrease the higher-end probabilities to another certain common value.
\end{theorem} 
The proof of Theorem \ref{thm:Qvalue_TV} and the closed-form solution are provided in Appendix \ref{pf:TV_solution}.  
The proof demonstrates that the solution follows a water-filling procedure to increase the low-probability components of $P^{i}$, and a reverse water-filling procedure to reduce the high-probability components, resembling the approach used in classical resource allocation problems.
If the $\delta$ constraint is loose, then the optimal $Q^{i}$ becomes the uniform distribution, i.e., the two common values become equal. Figure \ref{fig:tv_illus} provides a clear visualization of the solution in Theorem 2. 

\subsection{Adaptive $\delta$ Value Selection} \label{sec:delta_selection}

Let us denote the divergence threshold in each time $i$ as $\delta_i$. If \( \delta_i \) is set too large, the resulting adjustment to the probability distribution may lead to the selection of unusual tokens, negatively impacting the fluency of the stego-text. This issue is particularly noticeable when dealing with positions that have probability distributions with very low entropy values, i.e., most tokens have near-zero probability and the choices of tokens are almost deterministic. To address this issue, we need to choose \( \delta_i \) adaptively to the entropy $H(P^{i})$, i.e. $\delta_i = h(H(P^{i}))$. We take a straightforward approach in this work by setting $\delta_i = C\cdot H(P^{i})$ where $C$ is a constant. 

\section{Practical Considerations and Variations}

\subsection{Tokenization Errors} \label{sec:token_err}

LLM-based steganography relies on several assumptions, one of which is that Bob's tokenization process matches what was intended by Alice. This assumption is in fact quite subtle. The tokenizers in pre-trained LLMs guarantee that after detokenizing, the original text can be recovered; however, they do not guarantee to reproduce a unique sequence of tokens from any detokenized text. For example, Alice encodes the stego-text as \( \{ \text{``This''}, \text{``mount''}, \text{``ain''}, \text{``is''}, \text{``high''} \} \), forming ``This mountain is high.'' However, Bob may tokenize it as \( \{ \text{``This''}, \text{``mountain''}, \text{``is''}, \text{``high''} \} \), causing errors.
 In order words, the tokenizer merged ``mountain" into a single token rather than the two that the stego-text encoder intended. This issue exists in most of the previous LLM-based steganography approaches \citep{ziegler2019neural,shen2020near}, though only limited attention has been given to it \cite{10215094,yan2024near}. 

\begin{figure}[t]
    \includegraphics[width=\columnwidth]{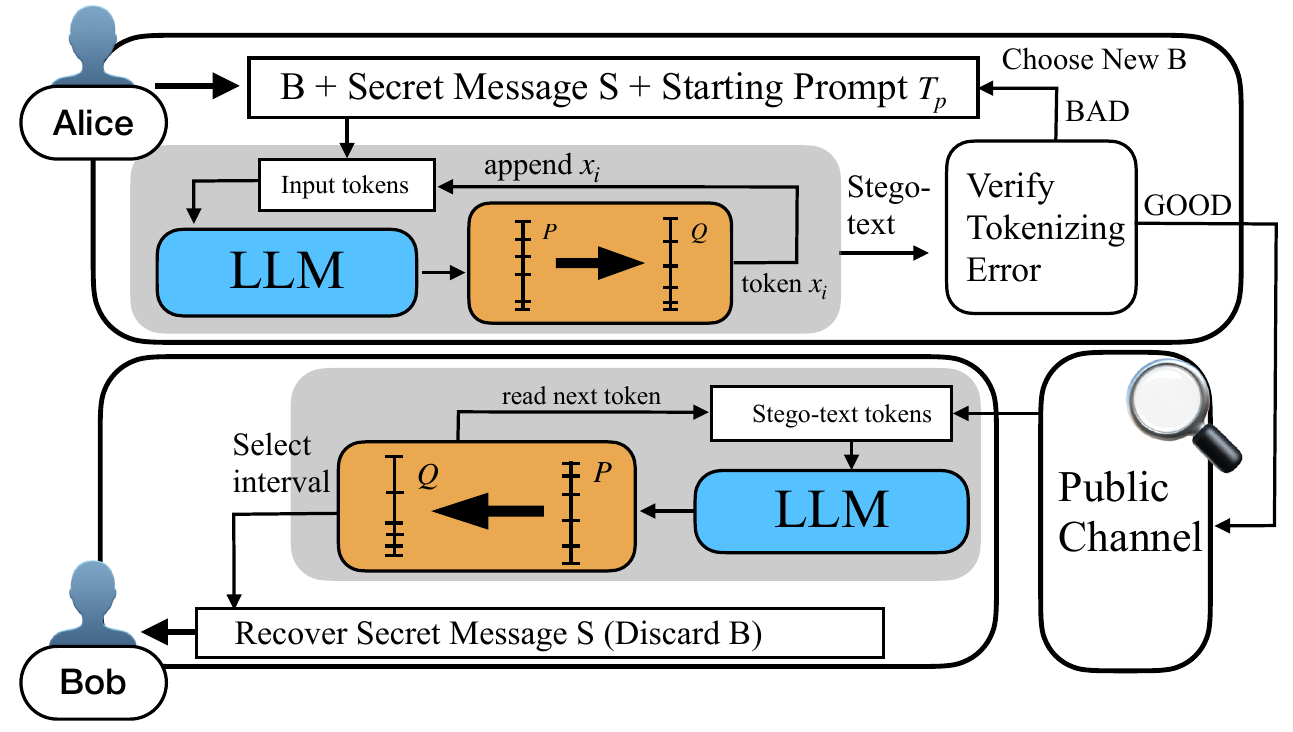}
    \caption{The OD-Stega approach}
    \vspace{-0.5cm}
    \label{fig:flowchart}
\end{figure}

This tokenization error leads Bob to decode a bit sequence different from the original secret bit sequence. 
Since LLMs are computationally demanding, it is too computationally expensive to enumerate such potential error cases to prevent such errors from occurring (see e.g. \cite{10215094,yan2024near} for efforts in this direction). Instead, we observe that tokenization errors are uncommon and the likelihood of such errors occurring is proportional to the length of the bit file. Moreover, Alice can in fact verify whether the stego-text can be correctly decoded by Bob since both have a copy of the same tokenizer. We therefore propose prepending a short sequence of additional $B$ bits to the bit sequence $S$ (a form of prompting). Alice then iterates among all $B$-bits combinations, and uses $f(T_p,(B,S))$ to produce the stego-text, until she verifies Bob can correctly decode. Bob simply discards the beginning $B$ bits after decoding. More experimental details to determine $B$ heuristically are given in Appendix \ref{sec:token}. The overall OD-Stega approach with this consideration is illustrated in Figure \ref{fig:flowchart}.

\subsection{Vocabulary Truncation}
\label{sec:trunc}
To reduce the computational complexity when the vocabulary set is large, a simple strategy is to truncate the vocabulary in the subsequent processing once a probability distribution has been generated. This strategy has been adopted in \cite{shen2020near}. To leverage our optimization formulation, we consider a two-stage process: first, we truncate the vocabulary, and second, we optimize the probability adjustment on the truncated vocabulary as discussed in the previous section. 
For this two-stage approach, we establish the KL divergence 
(and total variation) between the original distribution and the eventual optimized distribution on the truncated vocabulary, given below in Theorem \ref{thm:KL_additive}.

Let us make the two-stage strategy more precise. We first expand the zero-probability index set $\sA_i$ from $[N_{i}+1:N]$ to $[N_{\epsilon}+1:N]$, where $N_{\epsilon} = \min \{ n \mid \sum_{j = 1}^{n} P_j^{i} \geq 1 - \epsilon \}$. This leaves us with a total of $N_{\epsilon}$ variables. In addition, we define the re-normalized probability  $\hat{P}_j^{i}(\epsilon) = \frac{1}{1-\epsilon}P_j^{i}$, which we refer to as an $\epsilon$ cutoff probability of $P^{i}$. 
After the first stage, the variables in the optimization problem are reduced to $[Q_1^{i}, \cdots, Q_{N_{\epsilon}}^{i}]$.

The initial truncation phase creates the probability divergence $D(\hat{P}^{i}(\epsilon) || P^{i})$, and the second probability adjustment phase, as proposed in Section \ref{sec:opt_KL_strategy}, results in the divergence $D(Q^{i} || \hat{P}^{i}(\epsilon))$. 
The KL divergence does not satisfy the triangular inequality in general; however, in the specific case with a cutoff probability and its optimized counterpart, the following theorem demonstrates the additivity of KL divergences across these two stages.

\begin{theorem} \label{thm:KL_additive}
    Let $\hat{P}^{i}(\epsilon)$ be the $\epsilon$ cutoff probability distribution of $P^{i}$ and $Q^{i}$ be the solution of the optimization problem (\ref{eq:HPX})-(\ref{eq:px_equal_0}) with the constraint $D_{KL}(Q^{i}||\hat{P}^{i}(\epsilon)) \leq \hat{\delta}(\epsilon)$, then it holds that
    {\small \begin{align}
        D_{KL}(Q^{i} || {P}^{i}) & = D_{KL}(\hat{P}^{i}(\epsilon) || {P}^{i})  + D_{KL}(Q^{i} || \hat{P}^{i}(\epsilon)).
    \end{align}}
\end{theorem}
\begin{figure}[t]
\vspace{-0.3cm}
    \includegraphics[width=\columnwidth]{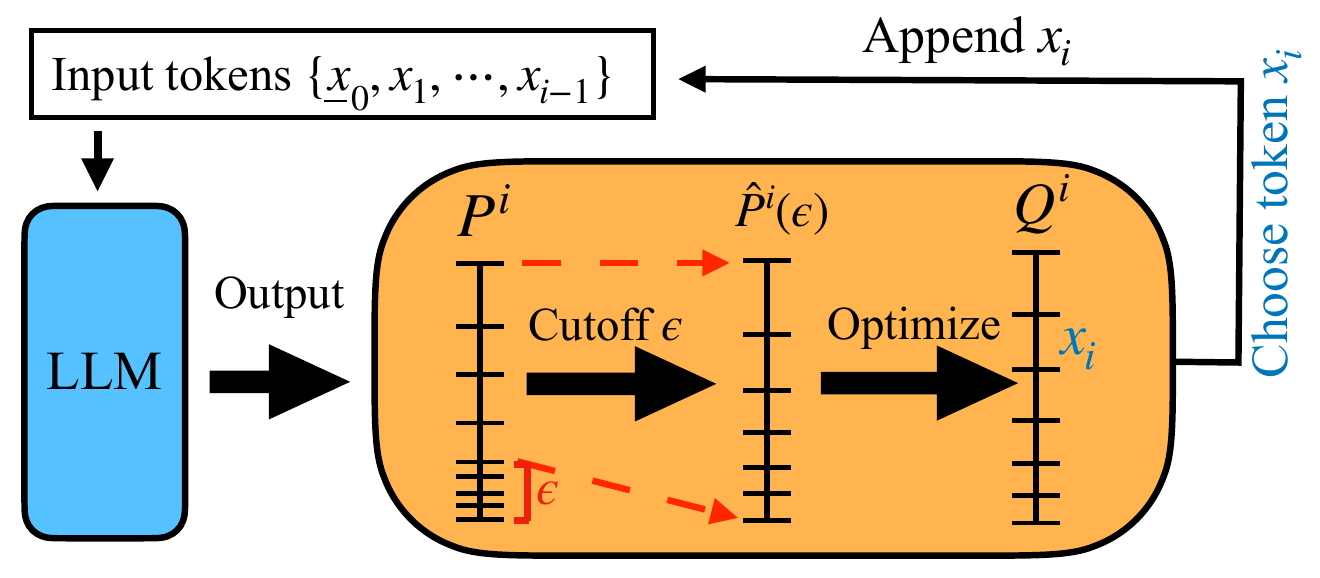}
    \vspace{-0.4cm}
    \caption{The two-stage design: Vocabulary truncation and distribution optimization}
    \label{fig:twostage}
    \vspace{-0.5cm}
\end{figure}
The proof can be found in Appendix \ref{sec:pf_thm3}.  
Given a total KL budget $\delta$, it is clear that we can determine $\hat{\delta}(\epsilon) = \delta -D_{KL}(\hat{P}^{i}(\epsilon) || P^{i})$.

For the case when TV is used in the divergence constraint, the triangular inequality allows us to set \textcolor{black}{$\hat{\delta}(\epsilon)=\delta - D_{TV}(\hat{P}^i(\epsilon) || P^{i}) = \delta - 2\epsilon$}, which guarantees the overall $D_{TV}(Q^{i} || {P}^{i})$ stays below $\delta$. Details are given in the Appendix \ref{pf:choice_of_deltahat_TV}.

\subsection{Variations and OD-Discop}

So far we have focused on using an AC decoder to directly drive the steganography encoder in a successive manner. Many existing AC-based methods can be naturally generalized to their relatively secure version in a straightforward manner, e.g., \citep{kaptchuk2021meteor, shen2020near, dai2019towards,ziegler2019neural, ivasenko2021information}; note that in the extreme case of without distribution optimization (i.e., $\delta_i=0$), these reduce to their original forms. More interestingly, our proposed approach even applies to techniques that do not directly rely on arithmetic coding. In the following, we discuss an adaptation of the Discop method \cite{ding2023discop}, which relies on distribution copies for encoding.

Discop \cite{ding2023discop} duplicates the probability distributions produced by an LLM, which the sampling distribution can follow exactly via pseudo-random numbers shared by Alice and Bob. The duplicated distributions are offset by different amounts, and the secret bits are encoded as the unique copy index in which the pseudo-number matches the sampled token. Though the exact mechanism becomes less transparent, it is straightforward to see that a highly non-uniform sampling distribution also induces low embedding utility in Discop. Therefore, adopting our proposed distribution-optimization technique can also improve embedding utility. To do so, we simply replace the LLM-produced distributions with our OD-adjusted version, prior to making the distribution copies, without impacting any other components of the coding pipeline. The corresponding experiments are given in the next section.

\section{Experimental Results}

\subsection{Experiment Setup} \label{sec:exp_setup}

We adopt the LLAMA2-7B pre-trained model \cite{touvron2023llama} as the underlying LLM in the experiment for OD-Stega, together with the SentencePiece tokenizer. We are aware that LLAMA3 \cite{grattafiori2024llama} and LLAMA4 \cite{meta_llama4_multimodal_2025} have become available; however, to evaluate the proposed technique, the larger models do not make any significant difference.
As mentioned earlier, we conducted experiments on OD-Discop, which combines Discop \cite{ding2023discop} with the proposed distribution optimization technique; in these experiments, the GPT2-XL pre-trained model with a capacity of 50,000 tokens is used. This was the LLM used in the original Discop, and we preserved it to avoid introducing unintentional confounders. The overall computational bottleneck is in the LLM and the additional computation of the proposed optimization is negligible, since closed-form solutions are available.

We performed experiments using a range of starting prompts on different topics of interest. Examples of topics include the Olympics, news, technology, and blogs, among others. The prompts usually have 10 to 20 words. In our two-stage optimization framework, we select a cutoff value \(\epsilon\) typically in the range \((0, 0.05]\), and also adjust the constant \(C \in [0, 0.1]\) to control the \(\delta_i\) values. 
Setting the cutoff $\epsilon$ at its maximum of $0.05$ results in the effective elimination of roughly 2000 lowest probability token choices for Llama2. By adjusting the range of $\delta_i$, 
we can assess how they impact the naturalness of the generated stego-texts and utilization. 
We referred to the stego-texts produced under the KL constraint as OD-KL, while those created under the TV constraint were called OD-TV.

The first evaluation metric is the embedding utilization, or equivalently the number of embedded bytes for a fixed number of generated stego-text tokens. The second quantity to evaluate is the naturalness (perceptability) of the generated stego-text, which is measured by three metrics: 1) The \textbf{KL Divergence} where a lower value implies better imperceptibility; 2) The detection rate using existing \textbf{steganalysis} techniques; 3) A perception evaluation using \textbf{GPT-4} as a human perception surrogate, where we simply ask GPT to determine whether the stego-text is written by human or not. 

Three different steganalysis techniques FCN \cite{yang2019fast}, SESY \cite{yang2021sesy}, and GS-Llama \cite{yang2024towards}, are chosen.
These techniques require training, and relevant details are given in Appendix \ref{sec:detection_setup}.

\subsection{Utilization-KL Tradeoff}\label{sec:expirementKL}

To study the embedding utilization performance, we keep the number of tokens in the stego-text fixed at 25, and evaluate how many secret bits can be embedded. We first verify that the constraining parameter $C$ indeed correlates linearly with the KL divergence of the sequence, and the results are shown in Appendix \ref{sec:KL_vs_C}. Equipped with this understanding, we focus on the study of the average number of bits that can be embedded vs. the parameter $C$ in the sequel. 

\begin{figure}[t]
    \includegraphics[width=\columnwidth]{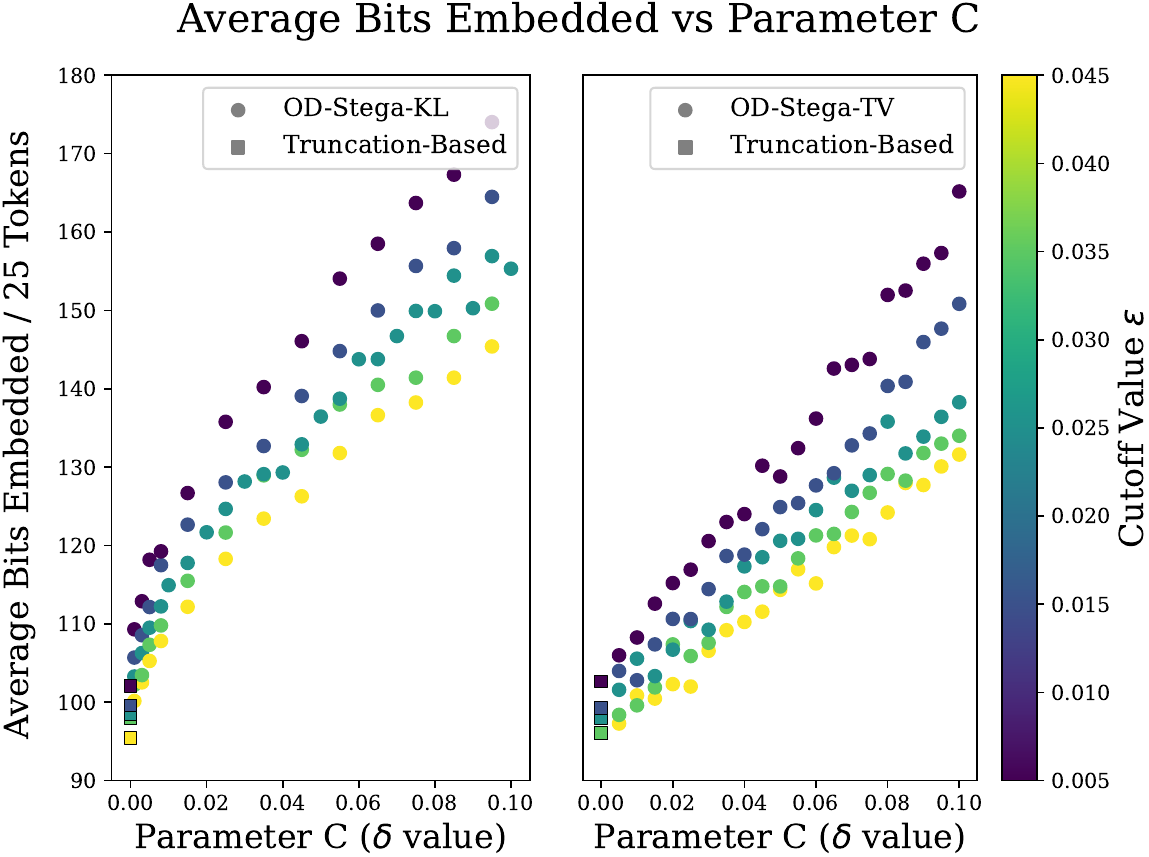}
    \caption{\textbf{OD-Stega.} Average bits embedded per 25 tokens (over 200 stego-texts) under KL and TV constraints, across parameter $C$ and cutoff $\epsilon$.}
    \vspace{-0.3cm}
    \label{fig:OD_Cvsbit_KL+TV}
\end{figure}

For OD-Stega, the parameter $C$ is varied from $0$ to $0.1$, and the truncation cutoff value parameter $\epsilon$ is varied from $0.005$ to $0.045$. For each $C$ and $\epsilon$, we average the numbers of bits embedded over 200 stego-texts to obtain the average. The results of OD-KL and OD-TV are illustrated in Figure \ref{fig:OD_Cvsbit_KL+TV}, where different colors indicate different truncation cutoff values, and the horizontal axis indicates the value of the parameter $C$.

\begin{figure}[t]
    \includegraphics[width=\columnwidth]{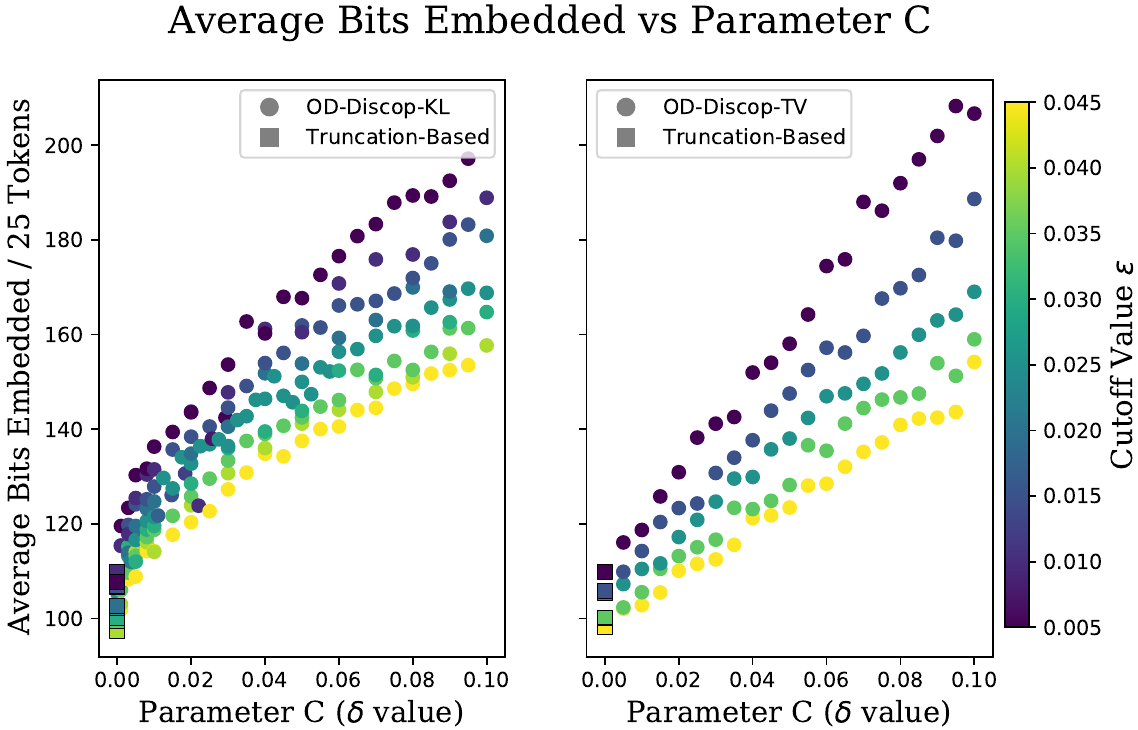}
    \caption{\textbf{OD-Discop.} Average bits embedded per 25 tokens (over 200 stego-texts) under KL and TV constraints, across parameter $C$ and cutoff $\epsilon$.}
    \vspace{-0.3cm}
    \label{fig:discop_Cvsbit}
\end{figure}

Observe first that the squares indicate the performance where only truncation is used without distribution optimization. It can be seen that embedding utilization decreases as the truncation value $\epsilon$ increases. Second, the points on the upper contour are produced with the lowest cutoff value $\epsilon=0.005$, consistent with the fact that higher cutoff reduces the embedding capacity. As $C$ increases (which corresponds to increasing $\delta_i$'s), the embedding utilization increases, initially more quickly and then roughly in a linear manner for OD-KL case. For OD-TV, the growth appears approximately linear throughout the entire range of $C$.

With $C=0.01$, which can be viewed as difficult to distinguish, i.e., when Eve is assumed to be a strong detector (see the next section for discussion on steganalysis techniques), the proposed method can increase embedding utilization by 20\%; If Eve is a weaker detector, then $C$ can be chosen at $C=0.05$, and the utilization can increase by 55\% in some cases.

In a similar manner, we evaluate the performance of OD-Discop, shown in Figure  \ref{fig:discop_Cvsbit}. 
We see a similar trend for OD-Discop, which confirms that the proposed approach indeed works as intended in Discop. The slight difference in the overall embedding utilization between Figure \ref{fig:OD_Cvsbit_KL+TV} and Figure \ref{fig:discop_Cvsbit} is mostly due to the completely different approaches in embedding the secret bits as well as the underlying LLMs, and should not be interpreted as one is better than the other.

\subsection{Steganalysis: Stego-Text Detection} \label{sec:detection_model}

Relatively secure steganography allows a weaker but controlled security degradation, and we now evaluate the behaviors of such degradation in OD-Stega and OD-Discop. Throughout this subsection, we fix the truncation cutoff parameter $\epsilon=0.025$. Note that current LLMs also usually include such a parameter, set at values ranging from $0.01$ to $0.05$.

Before performing tests on stego-texts, we first verify that setting $C=0$ in our OD-Stega approach indeed produces texts that are indistinguishable from LLM-generated texts without the arithmetic coding procedure at all. To do so, we utilize GS-Llama to obtain a classifier trained on a dataset with 500 texts from each class and then use it on a separate test dataset of 500 texts from each class. The classifier achieves roughly 50\% accuracy, implying that these two classes are indeed indistinguishable. Additional details can be found in Appendix \ref{sec:NSvsdelta0}.

As mentioned earlier, FCN, SESY, and GS-LLaMA are used to model Eve, who can have different levels of detection capabilities. We trained these detectors using OD-stega and OD-Discop stego-texts generated under both KL and TV constraints with various parameter $C$, combined with non-stego texts produced by the base language models (LLaMA2-7B and GPT2-XL). More details on the training procedure are given in Appendix \ref{sec:detection_setup}.

\begin{figure}[t]
    \includegraphics[width=\columnwidth]{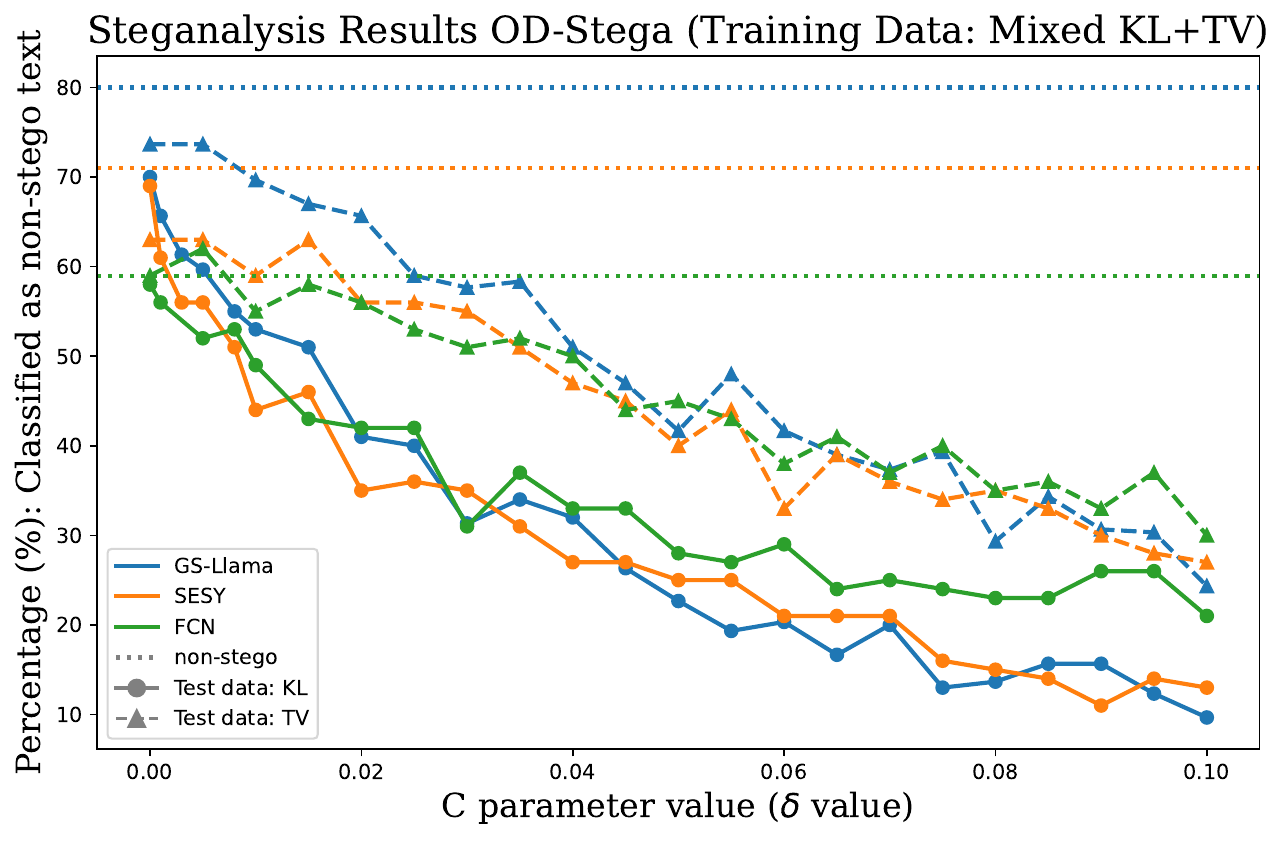}
    \caption{\textbf{Steganalysis on OD-stega coded texts.} The y-axis shows the fraction (\%) of stego-texts that the detector labels as \emph{non-stego}. Each data point is computed over 300 stego-texts for each value of parameter $C$.}
    \label{fig:OD_detection-ns}
\end{figure}

\begin{figure}[t]
    \includegraphics[width=\columnwidth]{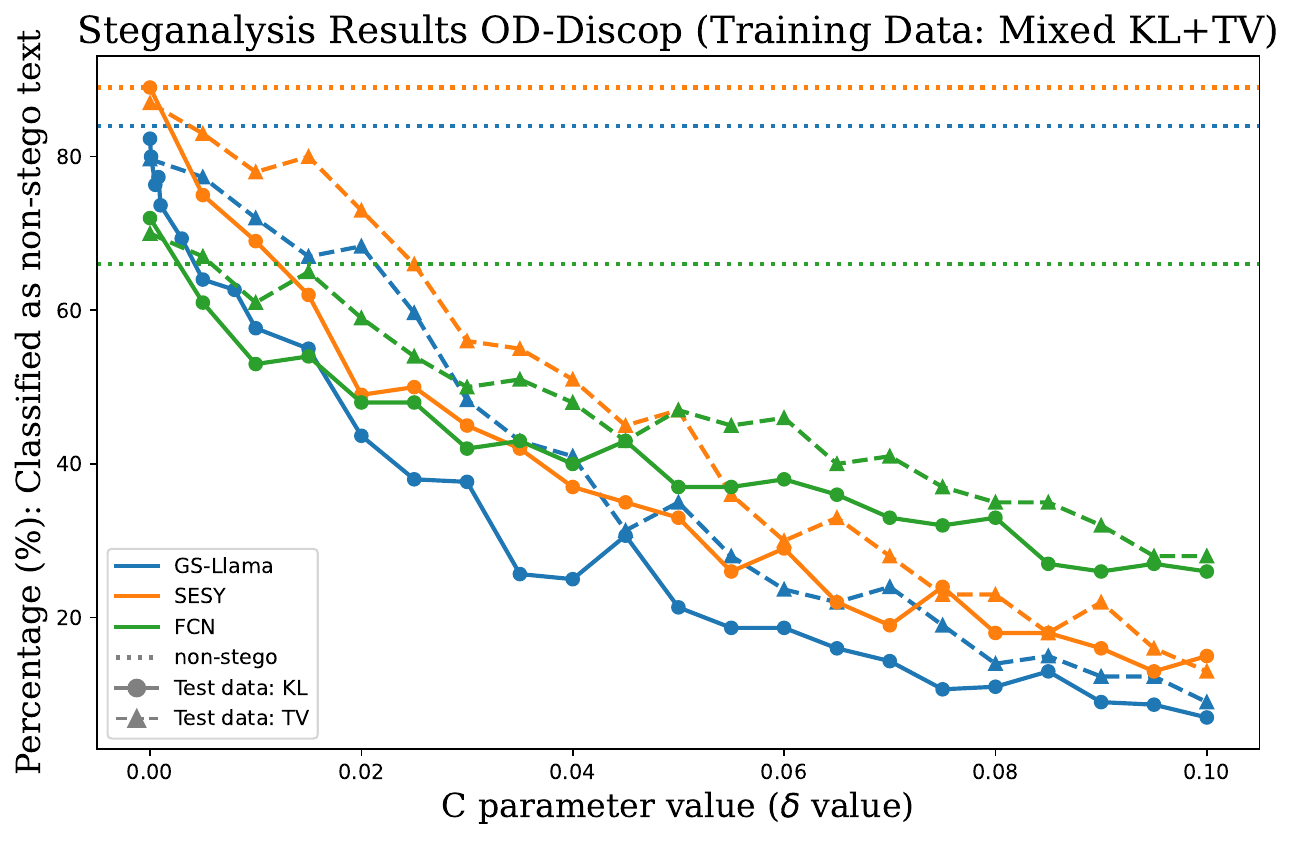}
    \caption{\textbf{Steganalysis on OD-Discop coded texts.} The y-axis shows the fraction (\%) of stego-texts that the detector labels as \emph{non-stego}. Each data point is computed over 300 stego-texts for each value of parameter $C$.}
    \label{fig:discop_detection_ns}
\end{figure}

For tests on OD-Stega and OD-Discop, the test dataset consists of 300 stego-texts for each $C$ value under KL and TV constraints, including the setting with $C=0$ and naturally generated texts without steganography coding. The results are shown in Figure \ref{fig:OD_detection-ns} and Figure \ref{fig:discop_detection_ns}. The circle markers correspond to the OD-KL method, while the triangular markers represent the OD-TV method. 
The horizontal dotted lines indicate the percentage non-stego texts that are correctly detected as non-stego, while the other lines indicate the percentage of stego-texts detected incorrectly as non-stego.   

In Figure \ref{fig:OD_detection-ns},
we see that as $C$ increases, the classifiers have a higher chance of classifying the stego-texts as so. In detecting OD-KL stego-texts, when \( C < 0.01 \), all three classifiers classify more than half of the stego-texts as non-stego. In the OD-TV case, the classifiers struggle to detect stego-texts when \( C < 0.05 \). Similar trends can be observed in the OD-Discop method, as shown in Figure~\ref{fig:discop_detection_ns}.
Additionally, as $C \rightarrow 0$, the percentage of stego-texts classified as stego-coded approaches that of LLM-generated texts.

In Figures \ref{fig:OD_detection-ns} and \ref{fig:discop_detection_ns}, the detectors may appear to perform similarly when considering only the stego-texts, however, FCN has a much lower percentage of correctly classifying natural texts, and similarly, SESY has an improved percentage of correctly classifying natural texts but not as high as GS-Llama. FCN is indeed considered the weakest model in the steganalysis literature that uses the least amount of computational resources. 
In Appendix \ref{sec:detection_results_acc}, Fig. \ref{fig:od-detection-acc} provides the overall detection accuracy vs. the parameter C, which illustrates this effect from another perspective. 
We additionally report steganalysis experiments for longer tokens (token length 100) in Appendix \ref{sec:detection_tk100}. The results exhibit similar trends to those for token length 25, with a slight shift for longer text sequences.

The behavior above confirms our early observation that the strength of the detector should be taken into account to achieve higher steganography embedding utilization. Additionally, given the linear relation between $C$ and KL-divergence in Fig. \ref{fig:OD_KLvsC}, we see that the latter is indeed a reliable measure to determine the imperceptibility threshold.

\subsection{GPT Evaluation} \label{sec:GPTevaluation}

We also use GPT-4 without fine-tuning to assess whether our stego-text appears natural and can escape detection by an untrained human-like evaluator. We instructed GPT to mimic a human evaluator to assess the text and determine if it was likely written by a human. In this experiment, we examined hundreds of generated stego-texts with GPT-4 under various parameters outlined in Section \ref{sec:expirementKL}. The OD-Stega result is shown in Figure \ref{fig:GPT4-eval-OD} (that for OD-Discop in Appendix \ref{sec:GPT_eval_results}), where the score indicates the percentage of texts rated as non-stego texts. The behavior is consistent with that in steganalysis: As the probability distribution gradually deviates further, more secret bits can be embedded at the expense of a gradually increased likelihood of attracting the evaluator's attention.

\begin{figure}[t]
    \vspace{-0.3cm}
    \includegraphics[width=\columnwidth]{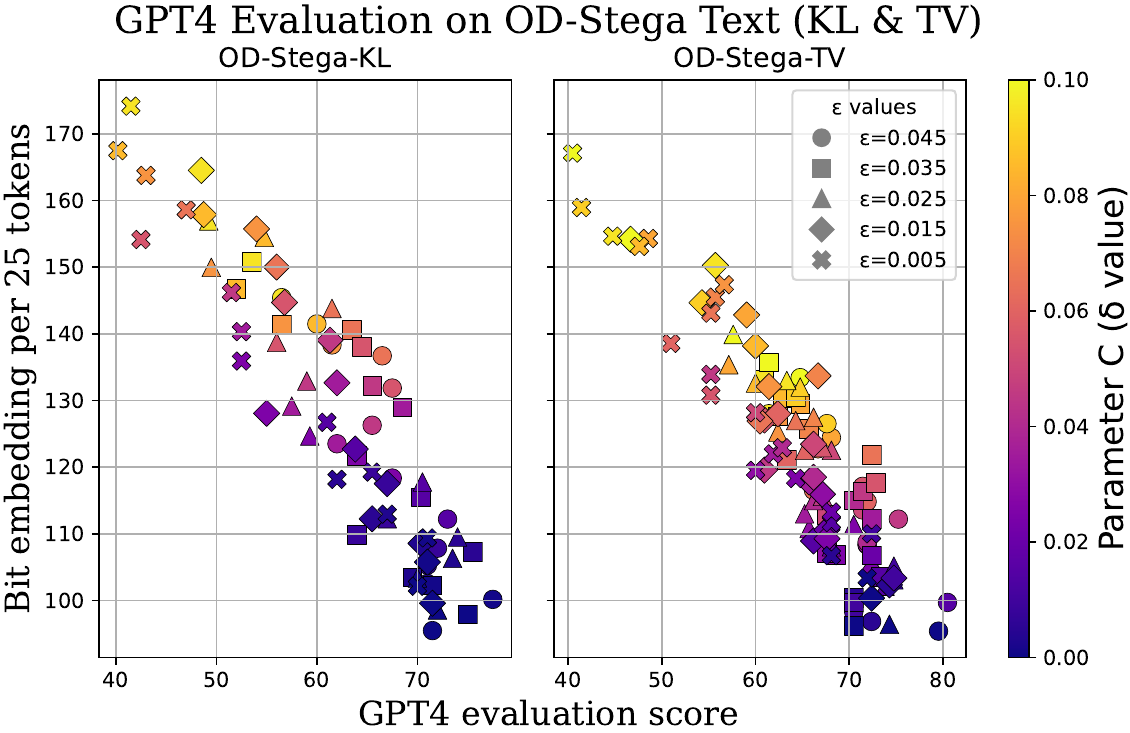}
    \caption{GPT-4 detection score on OD-Stega vs. average (over 200 texts) bits embedded per 25 tokens under KL and TV constraint.}
    \vspace{-0.3cm}
    \label{fig:GPT4-eval-OD}
\end{figure}

\subsection{Examples of Generated Stego-Texts}
We present several examples of stego-texts generated using OD-Stega-KL in Table~\ref{tab:example_text1}, each with different values of \( C \). The corresponding classification results from the SESY and FCN models are also shown. Here, NS denotes non-stego text and S denotes stego-text. As expected, larger values of \( C \) are more likely to induce unnaturalness and are generally easier for the classifier to detect as stego-text. However, this is not always the case. For example, the fourth example with $C=0.05$ is not classified as stego-text, while the example with $C=0.01$ is. It can also be seen that FCN makes a mistake in the third example, while SESY is correct.
Additional examples can be found in Appendix \ref{sec:moreexamples}.

\begin{table}[h]
\centering
\caption{Example stego-texts (OD-Stega-KL)}
\vspace{-0.3cm}
\renewcommand{\arraystretch}{1.2}
\scriptsize
\begin{tabularx}{\columnwidth}{l|X|l|l}
\toprule
Parameters                                                           & Prompt: Over the next few days, the weather will be & SESY & FCN\\ 
\midrule
\begin{tabular}[c]{@{}l@{}}$C = 0.01$\\ $\epsilon = 0.025$\end{tabular} & icy cold, with a temperature that could drop as low as -4 degrees on Christmas Eve. The Monterey and…  & NS & NS\\ \hline
\begin{tabular}[c]{@{}l@{}}$C = 0.05$\\ $\epsilon = 0.025$\end{tabular} & ATARSITY ADVENRTURE so if I didnt mention any of the places that I went … & S & S \\
\midrule
 & Prompt: Due to recent advances in technology,  & \\ 
\midrule
\begin{tabular}[c]{@{}l@{}}$C = 0.01$\\ $\epsilon = 0.025$\end{tabular} & 3D printing has revolutionized not only the operation of manufacturing sectors, but also current procedures. Homologation of... & S & NS  \\ \hline
\begin{tabular}[c]{@{}l@{}}$C = 0.05$\\ $\epsilon = 0.025$\end{tabular} & 3d printers are able to melt a plastic resin within the pores of a ceramic object… & NS & NS \\ 

\bottomrule
\end{tabularx}
\label{tab:example_text1}
\end{table}

\section{Conclusion}
We propose a technique to improve embedding utilization by formulating and solving a constrained optimization problem, resulting in the OD-Stega technique. We further address several practical issues, including tokenization errors, and vocabulary truncation. Moreover, variations such as OD-Discop were considered. We conducted extensive tests to show that the technique is indeed effective, and utilized both steganalysis tools and GPT to study the naturalness of the generated stega-texts.

\section{Limitations}

Relatively secure steganography can take advantage of the knowledge of the weak detector Eve's perceptibility, however, in this work, we did not consider the perceptibility question itself. Instead, we use the KL divergence as a surrogate to control it. An analogy is in image compression, where a compression method that can produce compressed images of different qualities may not specify the just noticeable difference (JND) value \cite{zhang2016just,liu2019deep}. The "imperceptible threshold" is clearly application-dependent and potentially individual-person-dependent, and therefore we treat it as a control lever in our work. Techniques to systematically design new perception neural networks to learn this threshold is beyond the scope of this work. 

The OD adjustment relies on optimizing each token individually using the conditional distribution causally. This process does not consider the impact that modifying the current token may have on the embedding capability for subsequent tokens. To take into account such long-term impact, one may need to consider a finite look-ahead approach, however, such an approach based on LLMs appears prohibitively expensive computationally \cite{huang2025relatively}. On the other hand, the KL divergence does decompose on the sequence level to the token level \cite{shen2020near}, therefore, the KL divergence on the sequence level can be well-controlled if it is well-controlled on the individual token level, although directly applying our individual-token-based optimization technique will not guarantee the optimality on the sequence level.

\bibliography{custom}

\newpage
\newpage
\begin{appendices}
\appendixpage
\section{Related Works}

Linguistic Steganography (LS) can be divided into two main areas: modification-based (cover-based) and generation-based (coverless). The modification-based approach conceals secret messages by altering the cover text through synonyms, syntactic changes, and word substitutions \citep{topkara2006hiding, chang2010linguistic, qi2013secure, chang2014practical}. In contrast, the generation-based approach creates stego-texts using methods like Markov chains \citep{dai2009bintext, dai2010text,moraldo2012approach} and deep learning techniques. With the advancement of generative language models, an increasing number of steganography research efforts now leverage neural networks to produce steganographic texts \citep{fang2017generating, yang2018rnn, ziegler2019neural,xiang2017novel,dai2019towards,zhang2021provably,shen2020near,kaptchuk2021meteor, ding2023discop,de2024perfectly}

\cite{fang2017generating}, for instance, explored a block-based methodology in which they designed a text generation model that first partitions the dictionary and allocates a specific code for each word. During the output stage, modified word-level LSTM neural network is utilized to choose words according to the encoded secret information. Their method organizes the vocabulary into subsets, the best word is chosen from a candidate pool based on the encoded bitstream at every generation step. \cite{yang2018rnn} presented a model that enhances text fluency and security in steganography by encoding each word dynamically based on its conditional probability distribution, employing both fixed-length coding (FLC) and variable-length coding (VLC). Through the use of structures like full binary trees or Huffman trees, this method enhances the naturalness and quality of generated texts while embedding hidden information more effectively.

\cite{ziegler2019neural} also utilized GPT-2 to create stego-texts, by proposing a linguistic steganography method that uses arithmetic coding with a pretrained neural language model. This method encodes secret messages by truncating the token distribution to the top $K$ most probable tokens at each generation step, thus minimizing the difference between the conditional probability distributions of steganographic and normal text, achieving close to optimal statistical security. Human evaluations were conducted to confirm that the generated text successfully deceived readers.

Building on Ziegler et al.'s arithmetic coding and truncating probability method, \cite{shen2020near} modified $K$ for each iteration, adjusting the conditional probability threshold with each new token. They claimed to select the smallest $K$ that still ensured near-imperceptibility. Additionally, they employed human evaluations to confirm their findings, demonstrating their method's effectiveness in deceiving eavesdroppers.

\cite{dai2019towards} employed GPT-2 for generating steganographic texts, crafting a novel steganographic mapping to embed secret messages and showcasing that effective mapping increases text security. They also proposed the patient-Huffman algorithm in such setting, which dynamically adjusts the embedding rate through the application of Kullback-Leibler divergence, enhancing both the quality and imperceptibility of steganographic texts. Their approach achieved near-imperceptibility, validated using total variation distance.

Recognizing the informal nature in the treatment of the security aspect of the methods in the studies from natural language processing community \cite{ziegler2019neural, dai2019towards, shen2020near}, the security research community further refined these methods to obtain provably secure protocols \citep{kaptchuk2021meteor,zhang2021provably,ding2023discop,de2024perfectly}. \cite{zhang2021provably} attempted to use grouping to match the granularity of probability to that of the secret message distribution granularity, however, their method is only perfectly secure when the natural language distribution allows such a grouping. Moreover, the grouping operation itelf also leads to a loss of embedding utilization. \cite{kaptchuk2021meteor} replaced the repeated secret key in \cite{ziegler2019neural} with pseudo-random generators, and showed that the resulting protocol is provably secure. However, the arithmetic coding component in \cite{kaptchuk2021meteor} is a reduced version from the full version, resulting in a slight loss in the embedding utilization. Instead of encrypting the original message and then using the generative model for steganography encoding, \cite{ding2023discop} combined the encryption step and the steganography encoding, resulting in another provably secure protocol. The work \cite{de2024perfectly} proposed a different approach to couple the message and the stego-text than using arithmetic coding directly.

In this paper, we present our encoding-decoding framework, drawing inspiration from \cite{ziegler2019neural} and \cite{shen2020near}. We observed that truncating a significant portion of the conditional probability from below leads to a reduction in bits embedded, which improves computational efficiency but reduces capacity. In fact, their approach for embedding long secret messages requires more computation in order to generate long stego-texts. To resolve this issue, we propose a novel method for adjusting the conditional probability to maximize the information embedded while maintaining near imperceptibility. 

\section{Proof of Remark \ref{remark:temp_scaling}} \label{sec:pf-of-remark1}
For each generation time $i$, if the LLM generated distribution $P^{i}$ results from the softmax of logits $Z^{i}$, the relation is shown as follows.
\begin{align}
    P^{i}_j = \frac{e^{Z^{i}_j}}{\sum_{k=1}^N e^{Z^{i}_k}}, ~ \forall j \in [1:N] \label{eq:pf-remark1-tempscale}
\end{align}
Applying temperature scaling to $P^{i}$ involve raising each probability in $P^{i}$ by the power $\frac{1}{T}$, followed by renormalization.
\begin{align}
    & \frac{{P^{i}_j}^{\frac{1}{T}}}{\sum_{n=1}^N {P^{i}_n}^{\frac{1}{T}}} =  \frac{\left( \frac{e^{Z^{i}_{j}}}{\sum^{N}_{k=1} e^{Z^{i}_{k}}} \right)^{\frac{1}{T} }  }{\sum^{N}_{n=1} \frac{e^{Z^{i}_{n}}}{\sum^{N}_{k=1} e^{Z^{i}_{k}}}^{\frac{1}{T} } } \label{eq:pf-remark1-1}\\
    = & \frac{\left( e^{Z^{i}_{j}}\right)^{\frac{1}{T} }  }{\left( \sum^{N}_{k=1} e^{Z^{i}_{k}}\right)^{\frac{1}{T} }  \sum^{N}_{n=1} \frac{\left( e^{Z^{i}_{n}}\right)^{\frac{1}{T} }  }{\left( \sum^{N}_{k=1} e^{Z^{i}_{k}}\right)^{\frac{1}{T} }  } } \\
    = & \frac{e^{\frac{Z^{i}_{j}}{T} }}{\sum^{N}_{n=1} e^{\frac{Z^{i}_{n}}{T} }}, ~ \forall j \in [1:N] \label{eq:pf-remark1-2}
\end{align}
The equivalence in (\ref{eq:pf-remark1-1}) follows from (\ref{eq:pf-remark1-tempscale}). Equations (\ref{eq:pf-remark1-1})-(\ref{eq:pf-remark1-2}) demonstrate that temperature scaling to $P^{i}$ matches the scaling to $Z^{i}$ by $\frac{1}{T}$ at the logit level, proving the remark.

\section{Proof of Theorem \ref{thm:Pxvalue_KL}}\label{pf:KL_solution}
The Lagrangian function of the problem (\ref{eq:HPX}) - (\ref{eq:px_equal_0}) with KL divergence constraint is 
\begin{align}
    \mathscr{L} & = \sum_{j=1}^{N_i} Q_j^{i} \log Q_j^{i} + u \left( \sum_{j=1}^{N_i} Q_j^{i} \log ( \frac{Q_j^{i}}{P_j^{i}}) - \delta \right) \notag \\
    & +\boldsymbol{\lambda}^{T}(-Q^{i}) + \omega \left( \sum_{j=1}^{N_i} Q_j^{i} -1 \right) \label{eq:lagrangian}
\end{align}
where $u, \boldsymbol{\lambda}, \omega$ are the Lagrangian multipliers of constraint (\ref{eq:KLconstraint}), (\ref{eq:postiveconstraint}) and (\ref{eq:sumeq1_constraint}), respectively. 
Then the KKT condition can be derived as follows:
 \begin{enumerate}
     \item Stationarity:
    \begin{align}
        \frac{\partial \mathscr{L}}{\partial Q_j^{i}} & = \log Q_j^{i} +1 + u \left( \log\frac{Q_j^{i}}{P_j^{i}} +1 \right) \notag \\
        & - \lambda_j + \omega = 0, \quad \forall j \in [1:N_i] \label{eq:KKT-stationarity}
    \end{align}

    \item Primal feasibility:
    \begin{align}
        \begin{cases}
         \sum_{j=1}^{N_i} Q_j^{i} \log \frac{Q_j^{i}}{P_j^{i}}  - \delta \leq 0\\ 
        Q_j^{i} \geq 0 , \quad \forall j \in [1:N_i]\\ 
        \sum_{j=1}^{N_i} Q_j^{i} -1 = 0 
        \end{cases} 
    \end{align}

    \item Dual feasibility:
    \begin{align}
        \begin{cases}u \geq 0 \\
        \lambda_j \geq 0, \quad \forall j \in [1:N_i] \\
        \end{cases} 
    \end{align}

    \item Complementary slackness:
    \begin{align}
        \begin{cases} u\left( \sum_{j=1}^{N_i} Q_j^{i} \log  \frac{Q_j^{i}}{P_j^{i}}  - \delta\right) =0  \\ 
        \lambda_{j }Q_j^{i} =0, ~ \forall j\in [1:N_i] \\ 
        \omega \left(\sum_{j=1}^{N_i} Q_j^{i} -1 \right)=0 \end{cases} 
    \end{align}
\end{enumerate}
Since the optimization problem is convex and clearly feasible, a solution to the KKT condition is also a global optimal solution. We claim the following is a solution to the KKT conditions:
\begin{enumerate}
    \item Primal variables: \\
    In case $\delta \in [0,\frac{1}{N_i} \sum_{j=1}^{N_i} \log(\frac{1}{N_i P_j^{i}}) ] $, from stationarity in (\ref{eq:KKT-stationarity}), 
    \begin{align}
        Q_j^{i} & = 2^{\frac{1}{1+u} \left( u\log P_j^{i} -1 + \lambda_j -u -\omega \right)} \\
        & = D{P_j^{i}} ^{\frac{u}{1+u}} , ~ \forall j \in [1:N_i]
    \end{align}
    where $D = 2^{\frac{-1 + \lambda_j -u -\omega}{1+u}}$ is a constant.\\
    Since $\sum_{j=1}^{N_i}Q_j^{i} = 1$, we can simply rewrite $Q_j^{i}$ in the form:
    \begin{align}
        Q_j^{i} & = \frac{{P_j^{i}}^{\frac{u}{1+u}}}{\sum_{j=1}^{N}{P_j^{i}}^{\frac{u}{1+u}}}, ~ \forall j \in [1:N_i]
    \end{align}
    In case $\delta > \frac{1}{N_i} \sum_{j=1}^{N_i} \log(\frac{1}{N_i P_j^{i}})$, we have
    \begin{align}
        Q_j^{i} & = \frac{1}{N_i}, ~ \forall j \in [1:N_i]
    \end{align}
    \item Dual variables:
    \begin{align}
        \begin{cases}
        u \begin{cases}
             \geq 0, ~ \delta \in [0,\frac{1}{N_i} \sum_{j=1}^{N_i} \log(\frac{1}{N_i P_j^{i}}) ] \\
             = 0, ~\delta > \frac{1}{N_i} \sum_{j=1}^{N_i} \log(\frac{1}{N_i P_j^{i}}) 
        \end{cases} \\
        \lambda_j = 0,  ~ \forall j \in [1:N_i] \\
        \omega = \frac{1}{1+u} \left( -1 + \log (\sum_{j =1}^{N_i} {P_j^{i}}^{\frac{u}{1+u}})\right) 
        \end{cases}
    \end{align}   
\end{enumerate}
It is straightforward to verify all the KKT conditions are satisfied, except the dual feasibility condition $u \geq 0$, which we prove in the next section.  

\section{Proof of $u \geq 0$ and Lemma \ref{lemma:ulargerthan0} First Part }\label{sec:pflemma_ulargerthan0} 

First, we note that 
\begin{align}
    & \lim_{u \rightarrow 0} D_{KL}(Q^{i} || P^{i}) = \frac{1}{N_i} \sum_{j=1}^{N_i} \log(\frac{1}{N_i P_j^{i}}) \\
    & \lim_{u \rightarrow \infty} D_{KL}(Q^{i} || P^{i}) = 0
\end{align}, because
\begin{align}
    \lim_{u \rightarrow 0} Q_j^{i} = \frac{1}{N_i}  \Rightarrow & \lim_{u \rightarrow 0}  D_{KL}(Q^{i} || P^{i}) \notag \\
    & = \sum_{j=1}^{N_i}\frac{1}{N_i} \log \left( \frac{\frac{1}{N_i}}{P_j^{i}}\right) \notag \\
    & = \frac{1}{N_i} \sum_{j=1}^{N_i} \log \left( \frac{1}{N_i P_j^{i}} \right) \\
    \lim_{u \rightarrow \infty} Q_j^{i}  = P_j^{i} 
     \Rightarrow & \lim_{u \rightarrow \infty}  D_{KL}(Q^{i} || P^{i})  \notag \\
     &= \sum_{j=1}^{N_i} P_j^{i} \log \left( \frac{P_j^{i}}{P_j^{i}} \right) = 0. 
\end{align}

Second, note that that $ D_{KL}(Q^{i} || P^{i})$ is continuous in $u \geq 0$. To see this, consider $P_j^{i}$ as the known distribution value, $Q_j^{i}$ is continuous in $u \geq 0$ because $\frac{u}{1+u}$ is continuous in $\mathbb{R} \setminus \{-1\}$. In addition, $Q_j^{i}$ will not be zero for all $j \in [1:N_i]$, which indicates that $\log (\frac{Q_j^{i}}{P_j^{i}})$ is continuous. Therefore, $ D_{KL}(Q^{i} || P^{i})$ is also continuous in $ u \geq 0$ since the function is a linear combination of continuous functions. 

 We will prove that $ D_{KL}(Q^{i} || P^{i})$ is non-increasing in $u$ for $u \geq 0$ in the next section. By the Intermediate-Value Theorem (IVT), it is clear that there exists a positive $u$ such that the KL divergence is equal to the given $\delta \in [0,\frac{1}{N_i} \sum_{j=1}^{N_i} \log(\frac{1}{N_i P_j^{i}}) ]$.

Here, we introduce the notation 
$\frac{1}{T} = \frac{u}{1+u}$
to simplify expressions and clarify the intuition behind the solution \(Q^{i}\). From this definition, it follows that $T = 1 + \frac{1}{u}$. 
Because there exists a positive \(u\) such that $D_{KL}\bigl(Q^{i} \,\|\, P^{i}\bigr) = \delta$ ,
we equivalently have a \(T \geq 1\) that satisfies this equality. This proves the first part of Lemma \ref{lemma:ulargerthan0}.

\section{Proof of Lemma \ref{lemma:ulargerthan0} Second Part}
Here we show that $ D_{KL}(Q^{i} || P^{i})$ is non-increasing in $u$ for $u \geq 0$, by analyzing the derivative as follows:  
\allowdisplaybreaks
\begin{align}
    & \frac{\partial  D_{KL}(Q^{i} || P^{i})}{\partial u} \notag \\
    &  = \frac{\partial}{\partial u}\left( \sum_{j=1}^{N_i} Q_j^{i} \log  \frac{Q_j^{i}}{P_j^{i}}\right) =\sum_{j=1}^{N_i} \frac{\partial}{\partial u}  \left(Q_j^{i} \log \frac{Q_j^{i}}{P_j^{i}}\right) \notag \\
    &  =  \sum_{j=1}^{N_i} \Biggl\{\frac{\partial}{\partial u}\left( \frac{{P_j^{i}}^{\frac{u}{1+u}}}{\sum_{k=1}^{N_i}{P^{i}_k}^{\frac{u}{1+u}}}\right)\log\left( \frac{{P_j^{i}}^{\frac{-1}{1+u}}}{\sum_{k=1}^{N_i}{P^{i}_k}^{\frac{u}{1+u}}}\right) \notag \\
    & \quad + \left( \frac{{P_j^{i}}^{\frac{u}{1+u}}}{\sum_{k=1}^{N_i}{P^{i}_k}^{\frac{u}{1+u}}}\right)\frac{\partial}{\partial u}\log\left( \frac{{P^{i}_j}^{\frac{-1}{1+u}}}{\sum_{k=1}^{N_i}{P^{i}_k}^{\frac{u}{1+u}}}\right) \Biggr\} \notag \\
    & = \sum_{j=1}^{N_i} \Biggl\{ \Biggl[ \frac{\left( \frac{1}{1+u}\right)^{2}  {P_j^{i}}^{\frac{u}{1+u}} }{\left( \sum_{k=1}^{N_i}{P_k^{i}}^{\frac{u}{1+u}}\right)^{2}} \Biggr] \notag \\
    &  \cdot \left( \sum_{k=1}^{N_i}{P_j^{i}}^{\frac{u}{1+u}} \log (\frac{P_j^{i}}{P^{i}_k}) \right) \cdot \log\left( \frac{{P_j^{i}}^{\frac{-1}{1+u}}}{\sum_{k=1}^{N_i}{P^{i}_k}^{\frac{u}{1+u}}}\right) \notag \\
    &  +\left( \frac{{P_j^{i}}^{\frac{u}{1+u}}}{\sum_{k=1}^{N_i}{P^{i}_k}^{\frac{u}{1+u}}}\right) \cdot    \frac{\left(\sum_{k=1}^{N_i}{P^{i}_k}^{\frac{u}{1+u}}\right) }{{P_j^{i}}^{\frac{-1}{1+u}} \left( \sum_{k=1}^{N_i}{P_k^{i}}^{\frac{u}{1+u}}\right)^{2}} \notag \\
    &   \cdot \left( \frac{1}{1+u}\right)^{2} {P_j^{i}}^{\frac{-1}{1+u}} \cdot \left( \sum_{k=1}^{N_i}{P^{i}_k}^{\frac{u}{1+u}} \log (\frac{P_j^{i}}{P^{i}_k}) \right) \Biggr\} \\
    & = \sum_{j=1}^{N_i} \left( \sum_{k=1}^{N_i}{P^{i}_k}^{\frac{u}{1+u}} \log (\frac{P_j^{i}}{P^{i}_k}) \right)\Biggl\{ \frac{\left( \frac{1}{1+u}\right)^{2} {{P_j^{i}}^{\frac{u}{1+u}}} }{\Gamma ^{2}} \cdot \notag \\
    &  \qquad \log\left( \frac{{P_j^{i}}^{\frac{-1}{1+u}}}{\Gamma }\right) +\left( \frac{{P_j^{i}}^{\frac{u}{1+u}} \Gamma  \left( \frac{1}{1+u}\right)^{2} }{\Gamma ^{3}}\right) \Biggr\} \\
    & = \frac{\left( \frac{1}{1+u}\right)^{2} }{\Gamma ^{2}} \sum_{j=1}^{N_i}\Biggl\{ {P_j^{i}}^{\frac{u}{1+u}} \left( \sum_{k=1}^{N_i}{P^{i}_k}^{\frac{u}{1+u}} \log (\frac{P_j^{i}}{P^{i}_k}) \right) \cdot \notag \\
    &  \qquad \qquad  \qquad   \left( \log\left( \frac{{P_j^{i}}^{\frac{-1}{1+u}}}{\Gamma }\right) +1 \right)\Biggr\} \\
    & =\frac{\left( \frac{1}{1+u}\right)^{2} }{\Gamma ^{2}} \sum_{j=1}^{N_i}\Biggl\{ \left( \sum_{k=1}^{N_i}({P_j^{i}}{P^{i}_k})^{\frac{u}{1+u}} \log (\frac{P_j^{i}}{P^{i}_k}) \right)\cdot \notag \\
    & \qquad \qquad  \qquad  \qquad  \qquad   \log\left( \frac{2{P_j^{i}}^{\frac{-1}{1+u}}}{\Gamma }\right) \Biggr\} \\
    & =\frac{\left( \frac{1}{1+u}\right)^{2} }{\Gamma ^{2}}  \sum_{j=1}^{N_i} \sum_{k=1}^{N_i} B_{jk}\log\left( \frac{2{P_j^{i}}^{\frac{-1}{1+u}}}{\Gamma }\right) \\
    & = \frac{\left( \frac{1}{1+u}\right)^{2} }{\Gamma ^{2}}  \sum_{\substack{j,k=1 \\ j \neq k \\ {P_j^{i}} \geq {P^{i}_k} }}^{N_i} B_{jk} \left( \log\left( \frac{2{P_j^{i}}^{\frac{-1}{1+u}}}{\Gamma }\right)- \right. \notag \\
    & \qquad  \qquad  \qquad  \qquad  \qquad  \left. \log\left( \frac{2{P^{i}_k}^{\frac{-1}{1+u}}}{\Gamma }\right) \right) \label{eq:Bjklargethan0}\\
    & = \frac{\left( \frac{1}{1+u}\right)^{2} }{\Gamma ^{2}}  \sum_{\substack{j,k=1 \\ j \neq k \\ {P_j^{i}} \geq {P^{i}_k} }}^{N_i} B_{jk}\left(\frac{-1}{1+u}\right)\log\left( \frac{{P_j^{i}}}{{P^{i}_k}} \right) \leq 0 \label{eq:KLdecreasing_proved}
\end{align}
where $\Gamma = \sum_{k=1}^{N_i}{P^{i}_k}^{\frac{u}{1+u}}$ and $B_{jk} = ({P_j^{i}}{P^{i}_k})^{\frac{u}{1+u}} \log \left( \frac{P_j^{i}}{P^{i}_k} \right)$. Eq.  (\ref{eq:Bjklargethan0}) follows from $B_{jk} = -B_{kj}, ~\forall j \neq k$ and $B_{jk} = 0,~\forall j = k$. The only negative term in (\ref{eq:KLdecreasing_proved}) is $\frac{-1}{1+u}$, since $B_{jk}$ and $\log(\frac{P_j^{i}}{P^{i}_k})$ are both positive in the case ${P_j^{i}} \geq {P^{i}_k} $. This proves the inequality. It follows that $ D_{KL}(Q^{i} || P^{i})$ is non-increasing in $u$ for $u \geq 0$.

Since the expression for $T$ is given by $T = 1 + \frac{1}{u}$, it follows that as the variable $u$ approaches $\infty$, $T$ approaches $1$. Conversely, when $u$ approaches to $0$, $T$ approaches infinity. Therefore, if $D_{KL}\bigl(Q^{i} \,\|\, P^{i}\bigr)$ behaves as a non-increasing function with respect to increasing values of $u$, the value will increase as $T$ itself increases. This proves the second part of the lemma \ref{lemma:ulargerthan0}.

\section{Proof of Theorem \ref{thm:Qvalue_TV}}

\label{pf:TV_solution}
In theorem \ref{thm:Qvalue_TV}, we address the optimization problem constrained by total variation in (\ref{eq:KLconstraint}), allowing us to reformulate it as:
\begin{align}
    \max_{Q_{j}^i} \quad & H(Q^{i}) \label{eq:TV_first_problem1}\\
    \text{subject to} \quad & D_{TV}(Q^i|| P^i) \leq \delta \\
    & Q_j^{i} \geq 0 \\
    & \sum_{j=1}^{N_i} Q_{j}^{i} = 1 \\
    & Q_{j}^{i} = 0 \quad \forall j \in \sA_i \label{eq:TV_first_problem5}
\end{align}

The total variation $D_{TV}(Q^i || P^i)$, as defined in (\ref{eq:def_TV}), is the norm 1 of the difference between $Q^i$ and $P^i$. Since all norms are convex, this problem remains a convex optimization problem and can be resolved by identifying the solution that meets all KKT conditions. 
\begin{align}
    D_{TV}(Q^i || P^i) = \sum_{j=1}^{N_i}  | Q_{j}^i- P_{j}^i | \label{eq:def_TV}
\end{align}
Clearly, the TV function is not differential at the point $Q_{j}^i = P_{j}^i$, resulting in the failure to achieve stationarity. Thus, we introduce new variables $b_j \geq 0 $ to this problem such that $| Q_{j}^i - P_{j}^i | \leq b_j$. The equivalent optimization problem (\ref{eq:TV_first_problem1})-(\ref{eq:TV_first_problem5}) corresponding to (\ref{eq:TV_second_problem1})-(\ref{eq:TV_second_problem6}) is now described as follows: 
\begin{align}
    \max_{Q_{j}^i, ~b_j ~\forall j \in [1:N_i] } \quad & H(Q^{i}) \label{eq:TV_second_problem1}\\
    \text{subject to} \quad & \sum_{j=1}^{N_i}b_j \leq \delta \label{eq:TV_second_problem2}\\
    & -b_j \leq  Q_j^i- P_j^i \leq b_j, ~\forall j \in [1:N_i]\\
    & Q_j^{i} \geq 0, ~\forall j \in [1:N_i] \\
    & \sum_{j=1}^{N_i} Q_{j}^{i} = 1 \\
    & b_j \geq 0, ~\forall j \in [1:N_i] \label{eq:TV_second_problem6} 
\end{align}
The solution to this problem is similar to the water-filling algorithm. To see this, we must recognize that the maximum entropy $Q^i$ occurs when it is a uniform distribution. However, modifications to $Q^i$ from $P^i$ are restricted by a given limit $\delta$. To allocate the value of each $b_j$ such that their sum equals $\delta$, the most straightforward method is to split $\delta$ in half, then water-fill the probabilities that are below average $\frac{1}{N_i}$, while decrease the probabilities that are above average, both with the total amount of $\frac{\delta}{2}$. A detailed closed-form solution is given below.

Assuming $P_j^{i}$ is sorted in descending order, let $S_H$ and $S_L$ be the index sets of tokens. Define $S_H = \{j \in [1:N_i] |P_{j}^i \geq \frac{1}{N_i} \}$ for high probability tokens, and $S_L =\{j\in [1:N_i]|P_{j}^i < \frac{1}{N_i} \}$ for low probability tokens.
The closed-form solution of optimal $Q_j^{i}$ in the optimization problem (\ref{eq:TV_second_problem1})-(\ref{eq:TV_second_problem6}) is described by different cases.  
\begin{enumerate}
    \item For $j \in S_L$, if $\frac{\delta}{2}$ satisfies the condition $(K_L+1)P_{{N_i-K_L-1}}^{i} - \sum_{k = N_i-K_L}^{N_i}P_{{k}}^{i} > \frac{\delta}{2} \geq K_LP_{{N_i-K_L}}^{i} - \sum_{k = N_i-K_L+1}^{N_i} P_{{k}}^{i}$ for some non-negative integer $K_L$ where $N_i-K_L \in S_L$, then 
    \begin{align}
        Q_{j}^i = \begin{cases}
           \Bar{P}_{Low}^{i}, ~ & j \in [N_i-K_L:N_i] \\
            P_{j}^i, ~ & j \in S_L \setminus [N_i-K_L:N_i] 
        \end{cases}.
    \end{align}
    where $\Bar{P}_{Low}^{i} = \frac{1}{K_L+1} \left( \frac{\delta}{2} + \sum_{k = N_i-K_L}^{N_i}P_{{k}}^{i} \right)$.
    Otherwise, if $\frac{\delta}{2} \geq \sum_{k \in S_L} (\frac{1}{N_i} - P_{{k}}^{i})$, then $Q_{j}^i = \frac{1}{N_i}, ~ \forall j \in S_L$.
    \item For $j \in S_H$, if $\frac{\delta}{2}$ satisfies the condition $ \sum_{k = 1}^{K_H+1}P_{{k}}^{i} -(K_H+1)P_{{K_H+2}}^{i} > \frac{\delta}{2} \geq  \sum_{k = 1}^{K_H} P_{{k}}^{i} - K_HP_{{K_H+1}}^{i}$ for some non-negative integer $K_H$ where $K_H+1 \in S_H$, then 
    \begin{align}
        Q_{j}^i = \begin{cases}
            \Bar{P}_{High}^{i}, ~ & j \in [1:K_H+1] \\
            P_{j}^i, ~ & j \in S_H \setminus [1: K_H+1] 
        \end{cases}.
    \end{align}
    where $\Bar{P}_{High}^{i} = \frac{1}{K_H+1} \left( \sum_{k = 1}^{K_H+1}P_{{k}}^{i} - \frac{\delta}{2} \right)$.
    Otherwise, if $\frac{\delta}{2} \geq \sum_{k \in S_H} (P_{{k}}^{i}-\frac{1}{N_i})$, then $Q_{j}^i = \frac{1}{N_i}, ~ \forall j \in S_H$.
\end{enumerate}
The intuitive illustration of the solution $Q^{i}$ is given in Figure \ref{fig:tv_illus}.

Given that this optimization problem is convex, the solution is obtained by solving the KKT conditions. The Lagrangian function for problem (\ref{eq:TV_second_problem1})-(\ref{eq:TV_second_problem6}) is 
\begin{align}
    \mathscr{L} & = \sum_{j=1}^{N_i} Q^{i}_{j} \log Q^{i}_{j} + u \left( \sum_{j=1}^{N_i} b_j - \delta \right) + \notag \\
    & \sum_{j=1}^{N_i} \alpha_j(-b_j - Q^{i}_{j} + P^{i}_{j}) + \notag \\
    & \sum_{j=1}^{N_i} \beta_j(Q^{i}_{j} - P^{i}_{j}-b_j)+ \omega(\sum_{j=1}^{N_i} Q^{i}_{j} -1)- \notag \\ 
    & \sum_{j=1}^{N_i}\lambda_j Q^{i}_{j} -\sum_{j=1}^{N_i}\eta_j b_j \label{eq:lagrangianTV} 
\end{align}
where $u, \alpha, \beta, \lambda, \omega, \eta$ are the Lagrangian multipliers of constraint in (\ref{eq:TV_second_problem2})-(\ref{eq:TV_second_problem6}).
Then the KKT condition can be derived as follows:
\begin{enumerate}
    \item stationarity:
    \begin{align}
    & \begin{cases}
        \frac{\partial \mathscr{L}}{\partial Q^{i}_{j}} & = \log Q^{i}_{j} +1 -\alpha_j + \beta_j + \omega - \lambda_j = 0 \\
        \frac{\partial \mathscr{L}}{\partial b_j} & = u - \alpha_j - \beta_j - \eta_j = 0 \label{eq:KKT-stationarityTV} \\
    \end{cases} \\
        \Rightarrow & \begin{cases}
            Q^{i}_{j} = e^{-1+\alpha_j - \beta_j - \omega +\lambda_j} \\
            u  = \alpha_j + \beta_j + \eta_j 
        \end{cases}
    \end{align}
    \item primal feasibility:
    \begin{align}
        \begin{cases}
         \sum_{j=1}^{N_i} b_j - \delta \leq 0 \\ 
        -b_j -Q^{i}_{j} + P_{j}^{i} \leq 0 , \quad \forall j \in [1:N_i]\\ 
        -b_j +Q^{i}_{j} - P_{j}^{i} \leq 0 , \quad \forall j \in [1:N_i]\\
        Q^{i}_{j} \geq 0, \quad \forall j \in [1:N_i]\\
        b_j \geq 0, \quad \forall j \in [1:N_i]\\
        \sum_{j=1}^{N_i} Q^{i}_{j} -1 = 0 
        \end{cases} 
    \end{align}

    \item dual feasibility:
    \begin{align}
        \begin{cases}u \geq 0 \\
        \alpha_j \geq 0, \quad \forall j \in [1:N_i] \\
        \beta_j \geq 0, \quad \forall j \in [1:N_i] \\
        \lambda_j \geq 0, \quad \forall j \in [1:N_i] \\
        \eta_j \geq 0, \quad \forall j \in [1:N_i] \\
        \end{cases} 
    \end{align}

    \item complementary slackness:
    \begin{align}
        \begin{cases} u\left( \sum_{j=1}^{N_i} b_j - \delta \right) =0  \\ 
        \alpha_j \left( -b_j -Q^{i}_{j} + P_{j}^{i} \right) =0  , \quad \forall j \in [1:N_i] \\
        \beta_j \left( -b_j +Q^{i}_{j} -P_{j}^{i} \right) =0  , \quad \forall j \in [1:N_i] \\
        \lambda_{j }Q^{i}_{j} =0, ~ \forall j\in [1:N_i] \\ 
        \eta_{j }b_j =0, ~ \forall j\in [1:N_i] \\
        \omega \left(\sum_{j=1}^{N_i}Q^{i}_{j} -1 \right)=0 \end{cases} 
    \end{align}
\end{enumerate}
We claim a solution to the KKT conditions as follows: \\
For $j \in S_L$,
 \begin{enumerate}
    \item Primal variables: \\
    The primal variables are the same as the formula stated in the Theorem. 
    If there exist some K where $N_i-K \in S_L$ and $\frac{\delta}{2} \in [KP_{{N_i-K}}^{i} - \sum_{k = N_i-K+1}^{N_i} P_{{k}}^{i},(K+1)P_{{N_i-K-1}}^{i} - \sum_{k = N_i-K}^{N_i}P_{{k}}^{i})$, then 
    \begin{align}
        Q^{i}_{j} & = \begin{cases}
        \frac{1}{K+1} \left( \frac{\delta}{2} + \sum_{k = N_i-K}^{N_i}P_{{k}}^{i} \right),\\ 
        \qquad \qquad \qquad ~~ j \in [N_i-K:N_i] \\
        P_{j}^i, ~ \quad \quad \quad j \in S_L \setminus [N_i-K:N_i]
        \end{cases} \\
        b_j & = \begin{cases}
        \frac{1}{K+1} \left( \frac{\delta}{2} + \sum_{k = N_i-K}^{N_i}P_{{k}}^{i} \right) - P_{{j}}^{i}, \\ ~ \qquad\qquad \qquad j \in [N_i-K:N_i] \\
        0, ~  j \in S_L \setminus [N_i-K:N_i]
        \end{cases} 
    \end{align}
    Otherwise, if $\frac{\delta}{2} \geq \sum_{k \in S_L} (\frac{1}{N_i} - P_{{k}}^{i})$, then $Q_{j}^i = \frac{1}{N_i}$ and $ b_j = \frac{1}{N_i} - P_{{j}}^{i} ,~ \forall j \in S_L$.
    \item Dual variables:\\
    Continue with the conditions above about $\frac{\delta}{2}$, for $\frac{\delta}{2} \in [KP_{{N_i-K}}^{i} - \sum_{k = N_i-k+1}^{N_i} P_{{k}}^{i},(K+1)P_{{N_i-K-1}}^{i} - \sum_{k = N_i-K}^{N_i}P_{{k}}^{i})$,
    \begin{align}
        \begin{cases}
            u & = -\log \left( \frac{1}{K+1} \left( \frac{\delta}{2} + \sum_{k = N_i-K}^{N_i}P_{{k}}^{i} \right) \right)\\
            & \qquad \qquad \qquad \qquad \qquad \quad -\log N_i\\
            \alpha_j & = 0, ~ \forall j \in S_L \\
            \beta_j & = \begin{cases}
                -\log \left( \frac{1}{K+1} \left( \frac{\delta}{2} + \sum_{k = N_i-K}^{N_i}P_{{k}}^{i} \right) \right)\\
                \qquad -\log N_i, ~~j \in [N_i-K:N_i] \\
                -\log P_{{j}}^{i} - \log N_i ~ ~, \\
                \qquad \qquad j \in S_L \setminus [N_i-K:N_i] \\
            \end{cases}\\
            \lambda_j & = 0, ~ \forall j \\
            \eta_j & = \begin{cases}
                0 ~ ,~~~ j \in [N_i-K:N_i] \\
                -\log \left( \frac{1}{K+1} \left( \frac{\delta}{2} + \sum_{k = N_i-K}^{N_i}P_{{k}}^{i} \right) \right)  \\
                 \quad + \log P_{{j}}^{i}, ~~j \in S_L \setminus [N_i-K:N_i] \\
            \end{cases}\\
            \omega & = \log N_i -1
        \end{cases}
    \end{align}
    The dual variables $u$ and all $\beta_j,\eta_j$s are positive because $\frac{1}{K+1} \left( \frac{\delta}{2} + \sum_{k = N_i-K}^{N_i}P_{{k}}^{i} \right) \leq P_{{l}}^{i} \leq \frac{1}{N_i}, ~ l \in S_L \setminus [N_i-K:N_i]$. \\
    Otherwise, if $\frac{\delta}{2} \geq \sum_{k \in S_L} (\frac{1}{N_i} - P_{{k}}^{i})$, the dual variables are 
    \begin{align}
        \begin{cases}
            u & = 0 \\
            \alpha_j & = \beta_j = 0, ~ \forall j \in S_L \\ 
            \lambda_j & = 0, ~ \forall j \\
            \eta_j & = 0, ~ \forall j \\
            \omega & = \log N_i -1
        \end{cases}
    \end{align}
    \end{enumerate}
For $j \in S_H$, 
\begin{enumerate}
    \item Primal Variables: \\
    Similarly, if there exists some K where $K+1 \in S_H$ and $\frac{\delta}{2} \in [\sum_{k = 1}^{K} P_{{k}}^{i} - KP_{{K+1}}^{i}, \sum_{k = 1}^{K+1}P_{{k}}^{i} -(K+1)P_{{K+2}}^{i})$, then
    \begin{align}
    Q_{j}^i & = \begin{cases}
        \frac{1}{K+1} ( \sum_{k = 1}^{K+1}P_{{k}}^{i} - \frac{\delta}{2} ), ~  j \in [1: K+1] \\
        P_{j}^i, ~  j \in S_H \setminus [1: K+1] 
    \end{cases} \\
    b_j & = \begin{cases}
    P_{{j}}^{i} - \frac{1}{K+1} \left( \frac{\delta}{2} + \sum_{k = N_i-K}^{N_i}P_{{k}}^{i} \right) , \\
    \qquad \qquad  \qquad \qquad j \in [1: K+1] \\
    0,  \quad j \in S_H \setminus [1: K+1] 
    \end{cases} 
    \end{align}
    Otherwise, if $\frac{\delta}{2} \geq \sum_{k \in S_H} (P_{{k}}^{i}-\frac{1}{N_i})$, then $Q_{j}^i = \frac{1}{N_i}$ and $b_j = P_{j}^i-\frac{1}{N_i}, ~ \forall j \in S_H$.
    \item Dual Variables: \\
     Continue with the conditions above about $\frac{\delta}{2}$, for $\frac{\delta}{2} \in [\sum_{k = 1}^{K} P_{{k}}^{i} - KP_{{K+1}}^{i}, \sum_{k = 1}^{K+1}P_{{k}}^{i} -(K+1)P_{{K+2}}^{i})$, 
     \begin{align}
        \begin{cases}
            u & = \log \left( \frac{1}{K+1} \left( \sum_{k = 1}^{K+1}P_{{k}}^{i} - \frac{\delta}{2} \right) \right) \\
            & \qquad\qquad\qquad\qquad\qquad+ \log N_i\\
            \alpha_j & = \begin{cases}
                \log \left( \frac{1}{K+1} \left( \sum_{k = 1}^{K+1}P_{{k}}^{i} - \frac{\delta}{2} \right) \right) \\
                 \qquad  \quad + \log N_i, ~~j \in [1:K+1] \\
                \log P_{{j}}^{i} + \log N_i, \\
                \qquad \qquad \quad j \in S_H \setminus [1:K+1] \\
            \end{cases}\\
            \beta_j & = 0, ~ \forall j \in S_H \\
            \lambda_j & = 0, ~ \forall j \in S_H \\
            \eta_j & = \begin{cases}
                0  , ~~~~j \in [1:K+1] \\
                \log \left( \frac{1}{K+1} \left( \sum_{k = 1}^{K+1}P_{{k}}^{i} - \frac{\delta}{2} \right) \right) \\
                \quad - \log P_{{j}}^{i} , j \in S_H \setminus [1:K+1] \\
            \end{cases}\\
            \omega & = \log N_i -1
        \end{cases}
    \end{align}
    The dual variables $u$ and all $\alpha_j,\eta_j$s are positive because $\frac{1}{K+1} \left( \sum_{k = 1}^{K+1}P_{{k}}^{i} - \frac{\delta}{2} \right) \geq P_{{l}}^{i} \geq \frac{1}{N_i}, ~ l \in S_H \setminus [1:K+1]$. \\
    Otherwise, if $\frac{\delta}{2} \geq \sum_{k \in S_H} (P_{{k}}^{i}-\frac{1}{N_i})$, then $Q_{j}^i = \frac{1}{N_i}, ~ \forall j \in S_H$.
    \begin{align}
        \begin{cases}
            u & = 0 \\
            \alpha_j & = \beta_j = 0, ~ \forall j \in S_H \\ 
            \lambda_j & = 0, ~ \forall j \\
            \eta_j & = 0, ~ \forall j \\
            \omega & = \log N_i -1
        \end{cases}
    \end{align}
\end{enumerate}
This completes the proof.

\section{Proof of Theorem \ref{thm:KL_additive}} \label{sec:pf_thm3}
The result in proof of Theorem \ref{thm:Pxvalue_KL} shows that the solution to the optimization problem with constraint $D_{KL}(Q^{i}||\hat{P}^{i}(\epsilon)) \leq \hat{\delta}(\epsilon)$ is: \\
If $\hat{\delta}(\epsilon) \in [ 0, \frac{ \sum_{j=1}^{N_\epsilon} \log(\frac{1}{N_{\epsilon}{\hat{P}_j^{i}}})}{N_{\epsilon}}]$, 
\begin{align}
    Q_j^{i} = \frac{{{\hat{P}_j^{i}}(\epsilon)}^{\frac{\hat{u}(\epsilon)}{1+\hat{u}(\epsilon)}}}{\sum_{j=1}^{N_{\epsilon}}{\hat{P}_j^{i}(\epsilon)}^{\frac{\hat{u}(\epsilon)}{1+\hat{u}(\epsilon)}}}
\end{align}

Otherwise,
\begin{align}
    Q_j^{i} = \frac{1}{N_{\epsilon}},
\end{align}
for all $j \in [1:N_\epsilon]$ and for some positive $\hat{u}(\epsilon)$.

In addition, Lemma \ref{lemma:ulargerthan0} states that when  $\hat{\delta}(\epsilon)\in [ 0, \frac{1}{N_{\epsilon}} \sum_{j=1}^{N_\epsilon} \log(\frac{1}{N_{\epsilon}{\hat{P}_j^{i}}})]$, the obtained solution ensures that the KL divergence $D_{KL}(Q^{i} || \hat{P}^{i}(\epsilon))$ is equal to the given constraint $ \hat{\delta}(\epsilon)$.
We first calculate the KL divergence of the truncation process, which is the distance between $P^{i}$ and $\hat{P}^{i}(\epsilon)$.
\begin{align}
   &  D_{KL}(\hat{P}^{i}(\epsilon) || P^{i})  = \sum_{j=1}^{N_{\epsilon}}\hat{P}_j^{i}(\epsilon) \log\left(\frac{\hat{P}_j^{i}(\epsilon)}{P_j^{i}}\right) \notag \\
    & = \sum_{j=1}^{N_{\epsilon}} \frac{1}{1-\epsilon} P_j^{i} \log\left(\frac{\frac{1}{1-\epsilon}P_j^{i}}{P_j^{i}}\right) \notag \\
    & = \frac{1}{1-\epsilon} \log \left(\frac{1}{1-\epsilon}\right) \sum_{j=1}^{N_{\epsilon}}P_j^{i}  = - \log ({1-\epsilon}) \label{eq:cutoff_delta}
\end{align}

In the following, we show that in this case, the KL divergence is additive, which means that the divergence between $Q^{i}$ and ${P}^{i}$ is the sum of the divergence between $Q^{i}$ and $\hat{P}^{i}(\epsilon)$ and between $\hat{P}^{i}(\epsilon)$ and ${P}^{i}$.
\begin{align}
    & \hat{\delta}(\epsilon)  = \sum_{j=1}^{N_{\epsilon}} {Q}^{i}_j \log \left( \frac{{Q}^{i}_j}{\hat{P}_j^{i}(\epsilon)}\right)\\
    & = \sum_{j=1}^{N_{\epsilon}} \left( \frac{{{\hat{P}_j^{i}}(\epsilon)}^{\frac{\hat{u}(\epsilon)}{1+\hat{u}(\epsilon)}}}{\sum_{k=1}^{N_{\epsilon}}{\hat{P}_k^{i}(\epsilon)}^{\frac{\hat{u}(\epsilon)}{1+\hat{u}(\epsilon)}}} \right) \cdot \notag \\
    & \qquad  \qquad  \qquad  \log \left( \frac{{{\hat{P}_j^{i}(\epsilon)}}^{\frac{\hat{u}(\epsilon)}{1+\hat{u}(\epsilon)}}}{\hat{P}_j^{i}(\epsilon) \sum_{k=1}^{N_{\epsilon}}{\hat{P}_k^{i}(\epsilon)}^{\frac{\hat{u}(\epsilon)}{1+\hat{u}(\epsilon)}}} \right) \\
    & = \sum_{j=1}^{N_{\epsilon}} \left( \frac{{{P_j^{i}}}^{\frac{\hat{u}(\epsilon)}{1+\hat{u}(\epsilon)}}}{\sum_{k=1}^{N_{\epsilon}}{P_k^{i}}^{\frac{\hat{u}(\epsilon)}{1+\hat{u}(\epsilon)}}} \right)  \cdot \notag \\
    & \qquad  \qquad  \qquad  \log \left( \frac{(1-\epsilon ){{P_j^{i}}}^{\frac{\hat{u}(\epsilon)}{1+\hat{u}(\epsilon)}}}{P_j^{i} \sum_{k=1}^{N_{\epsilon}}{P_k^{i}}^{\frac{\hat{u}(\epsilon)}{1+\hat{u}(\epsilon)}}} \right) \\ 
    & = \sum_{j=1}^{N_{\epsilon}} \left( \frac{{{P_j^{i}}}^{\frac{\hat{u}(\epsilon)}{1+\hat{u}(\epsilon)}}}{\hat{\Gamma}(\epsilon)} \right) \log \left( \frac{(1-\epsilon){{P_j^{i}}}^{\frac{-1}{1+\hat{u}(\epsilon)}}}{\hat{\Gamma}(\epsilon)} \right) \label{eq:tildeT}\\
    & = \sum_{j=1}^{N_{\epsilon}} \left( \frac{{{P_j^{i}}}^{\frac{\hat{u}(\epsilon)}{1+\hat{u}(\epsilon)}}}{\hat{\Gamma}(\epsilon)} \right)  \Bigg\{ \log (1-\epsilon) +  \notag \\
    &  \qquad  \qquad  \qquad  \qquad  \log \left( \frac{{{P_j^{i}}}^{\frac{-1}{1+\hat{u}(\epsilon)}}}{\hat{\Gamma}(\epsilon)} \right)  \Bigg\} \\
    & = \sum_{j=1}^{N_{\epsilon}} \left( \frac{{{P_j^{i}}}^{\frac{\hat{u}(\epsilon)}{1+\hat{u}(\epsilon)}}}{\hat{\Gamma}(\epsilon)} \right)\log \left( \frac{{{P_j^{i}}}^{\frac{-1}{1+\hat{u}(\epsilon)}}}{\hat{\Gamma}(\epsilon)} \right)+ \notag \\
    & \qquad  \qquad  \qquad  \qquad  \frac{\log (1-\epsilon)}{\hat{\Gamma}(\epsilon)}\sum_{j=1}^{N_{\epsilon}}{{P_j^{i}}}^{\frac{\hat{u}(\epsilon)}{1+\hat{u}(\epsilon)}} \\
    & = D_{KL}(Q^{i} || {P}^{i}) + \log (1-\epsilon) \\
    & = D_{KL}(Q^{i} || {P}^{i}) - D_{KL}(\hat{P}^{i}(\epsilon) || {P}^{i}) \label{eq:KLadditive_step8}\\
    & \Rightarrow ~  D_{KL}(Q^{i} || {P}^{i}) \notag \\
& = D_{KL}(\hat{P}^{i}(\epsilon) || {P}^{i}) + D_{KL}(Q^{i} || \hat{P}^{i}(\epsilon))
\end{align}
where $\hat{\Gamma}(\epsilon) = \sum_{j=1}^{N_{\epsilon}}{{P}^{i}_j}^{\frac{\hat{u}(\epsilon)}{1+\hat{u}(\epsilon)}}$ in (\ref{eq:tildeT}) and equation (\ref{eq:KLadditive_step8}) follows from (\ref{eq:cutoff_delta}).\\

\section{Choice of $\hat{\delta}(\epsilon)$ Under TV Constraint} \label{pf:choice_of_deltahat_TV}
Under TV constraint, the optimal solution $Q^i$ to this optimization problem is given in theorem \ref{thm:Qvalue_TV}. By adopting a probability cutoff before adjustments, the TV distance upper bound between $P^{i}$ and $Q^{i}$ becomes the sum of the cutoff and adjustment TV distances, due to the triangular inequality.
\begin{align}
    D_{TV}(Q^{i}||P^{i}) & \leq D_{TV}(\hat{P}^{i}(\epsilon)||P^{i}) \notag \\
    & \qquad \quad + D_{TV}(Q^{i}||\hat{P}^{i}(\epsilon)) \label{eq:Tv_cutoff_delta_UB} 
\end{align}

Calculating the TV distance between the original probability and the reduced probability is straightforward:
\begin{align}
    & D_{TV}(\hat{P}^{i}(\epsilon)||P^{i})\notag\\
    & =  \sum_{j = 1}^{N_{\epsilon}} |\frac{1}{1- \varepsilon} P_j^{i} - P_j^{i}| + \sum_{j = N_{\epsilon}+1}^{N} |P_j^{i} - 0|  \\  
    & = \frac{\varepsilon}{1- \varepsilon} \sum_{j = 1}^{N_{\epsilon}} P_{j}^{i} + \varepsilon = 2\varepsilon. \label{eq:TV_cutoff_value}
\end{align}
Assuming the cutoff tokens $[N_{\epsilon}+1:N]$ have probabilities summing up to $\epsilon$, the rest of the probabilities from $[1:N_{\epsilon}]$ sum to $1-\epsilon$, leading to the equality in equation (\ref{eq:TV_cutoff_value}).
Therefore, the equation (\ref{eq:Tv_cutoff_delta_UB}) can be further derived as: 
\begin{align}
    & D_{TV}(Q^{i}||P^{i}) \leq 2\varepsilon + \hat{\delta}(\epsilon), \notag \\
    & \qquad \qquad ~ \text{for} ~ \hat{\delta}(\epsilon) \in [0, \sum_{j=1}^{N_\epsilon} |\hat{P}_j^{i}(\epsilon) - \frac{1}{N_\epsilon}|] 
\end{align}
However, this upper bound is not a strict upper bound. In fact, the actual value of the total variation between $Q^{i},P^{i}$ may be slightly lower than this upper bound, given that the TV divergence is not additive, this differs from the conclusion in Theorem \ref{thm:KL_additive}, where we found that the KL divergence is additive in the particular case.

\section{KL divergence vs. Parameter C} \label{sec:KL_vs_C}

\begin{figure}[]
    \includegraphics[width=\columnwidth]{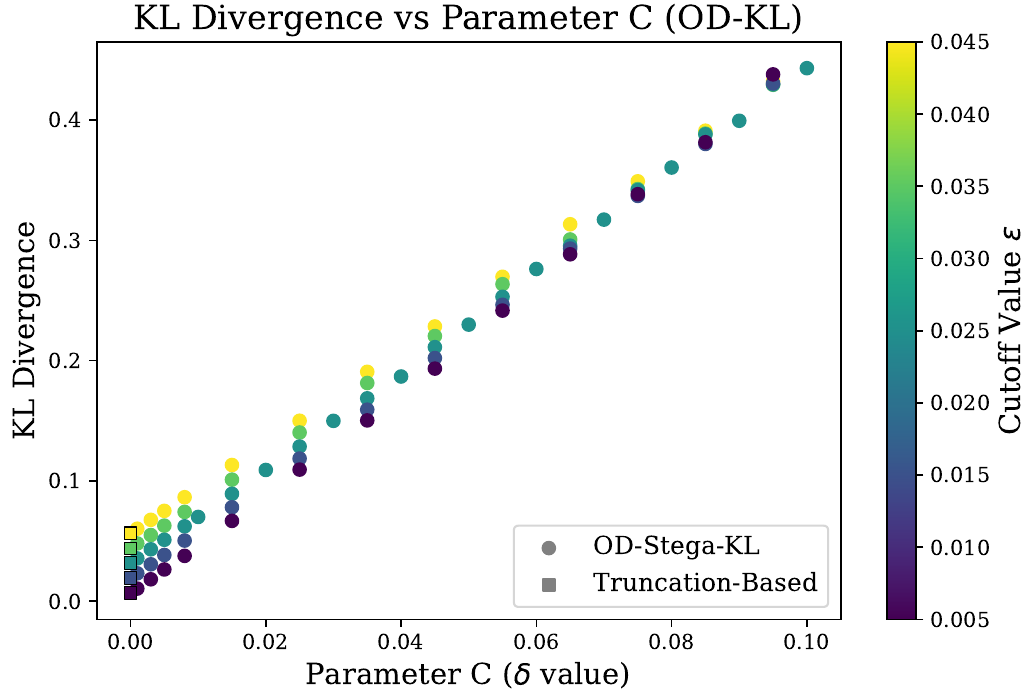}
    \caption{KL-divergence vs. parameter C  (OD-KL). }
    \vspace{-0.3cm}
    \label{fig:OD_KLvsC}
\end{figure}

\begin{figure}[h]
    \includegraphics[width=\columnwidth]{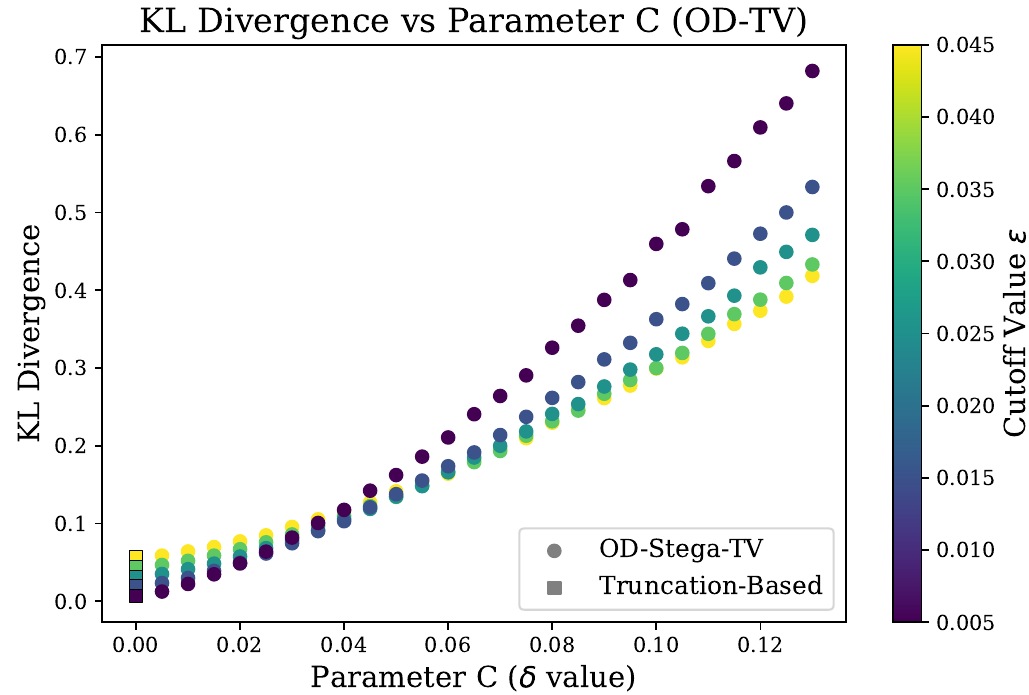}
    \caption{KL-divergence vs. parameter C (OD-TV). }
    \vspace{-0.3cm}
    \label{fig:ODTV_KLvsC}
\end{figure}

In this section, we demonstrated that our adjustment parameter $C$ is strongly correlated with the KL divergence between stego-text and natural text. As shown in Figure \ref{fig:OD_KLvsC}, these two values exhibit an almost linear relationship, indicating that as $C$ increases, the distortion relative to the original probability distribution given by the LLM also increases. This result is intuitive since we define $\delta_i = C \cdot H(P^i)$, meaning that a larger $C$ directly leads to a larger $\delta_i$, which is a larger adjustment. 
Figure \ref{fig:ODTV_KLvsC} gives a comparable plot for OD-TV. Despite the non-linear relationship between KL divergence and parameter C here, the figure illustrates that C remains effective in controlling the probability divergence. 


\section{OD-KL v.s. Temperature Adjsutment}
\begin{figure}[h]
    \includegraphics[width=\columnwidth]{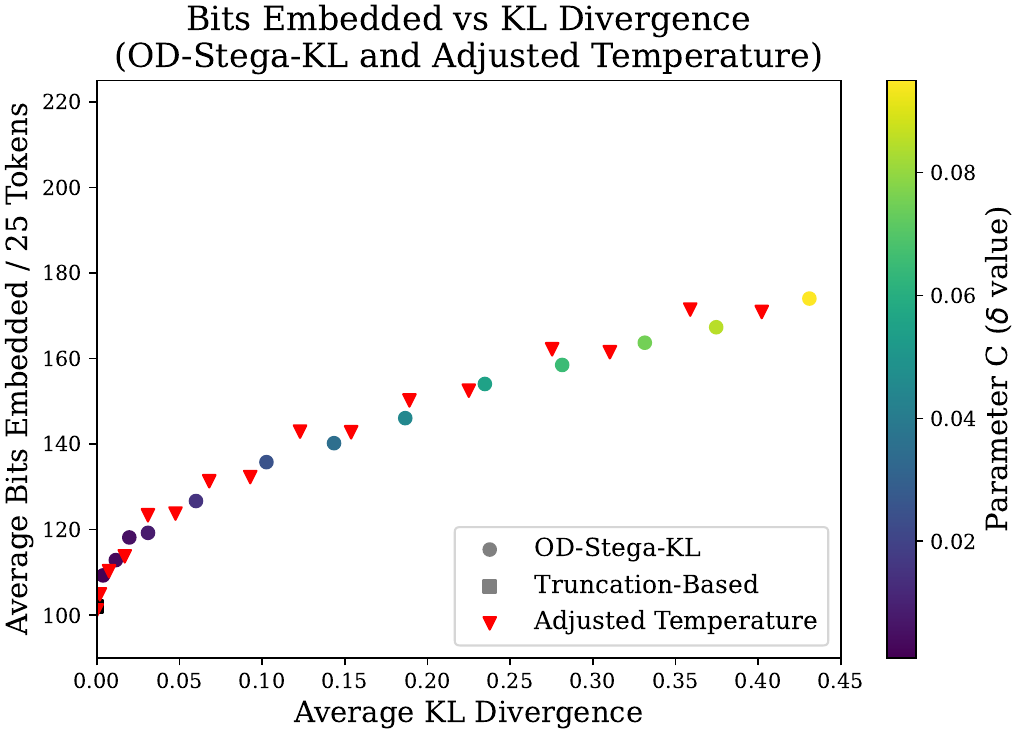}
    \caption{OD-KL vs. Adjusted Temperature}
    \vspace{-0.3cm}
    \label{fig:KL_vs_temp}
\end{figure}

This section presents an experiment comparing OD-KL adjustments with direct temperature scaling in the LLM. In Figure \ref{fig:KL_vs_temp}, the x-axis represents the KL divergence between the adjusted distribution and the original LLM distribution, while the y-axis indicates the number of bits that can be embedded into the stego-text, which corresponds to the entropy. The red triangles denote the results obtained by tuning the temperature parameter, whereas the OD-KL adjustment is computed using our closed-form expression provided in Theorem \ref{thm:Pxvalue_KL}. We observe that the two methods yield identical results, which confirms our claim in Remark \ref{remark:temp_scaling} that OD-KL adjustment is equivalent to temperature scaling at the logits level.

\section{Details for Steganalysis Models} \label{sec:detection_setup}

For the three steganalysis techniques, the models were trained using stego-texts generated with the parameter \( C \) ranging from 0.01 to 0.1 (in step 0.01) and with \( \epsilon = 0.025 \), combining generations under both KL and total variation (TV) constraints. These were mixed with naturally generated texts without any steganographic encoding.
For each of the 10 distinct $C$ values, 200 stego-texts are included in the training set, alongside 4000 naturally generated texts (non-stego) generated from Llama2-7B and GPT2-XL. For GS-Llama, we applied LoRA to fine-tune Llama-3.1-8B-Instruct, optimizing for efficiency while maintaining performance. The model underwent 10 epochs of training (24 GPU hours) using one NVIDIA H100 GPU (80GB memory), a learning rate of $1\times 10^{-5}$, FP16 precision, a sequence length of 128, and the AdamW optimizer, without implementing a validation split to leverage the data fully. For FCN, training was conducted over 300 epochs with a learning rate of \( 5 \times 10^{-4} \). For SESY, training was performed over 10 epochs using the AdamW optimizer with a learning rate of \( 5 \times 10^{-5} \). The model was trained for 10 epochs, which took approximately 8 GPU hours on a single NVIDIA A100 GPU with 40\,GB of memory. Other parameters were kept as the default values in FCN \cite{yang2019fast}, SESY \cite{yang2021sesy}, and GS-Llama \cite{yang2024towards}.

\section{Alternative Plots for Steganalysis Results} \label{sec:detection_results_acc}

\begin{figure}[ht]
    \centering
    \includegraphics[width=\linewidth]{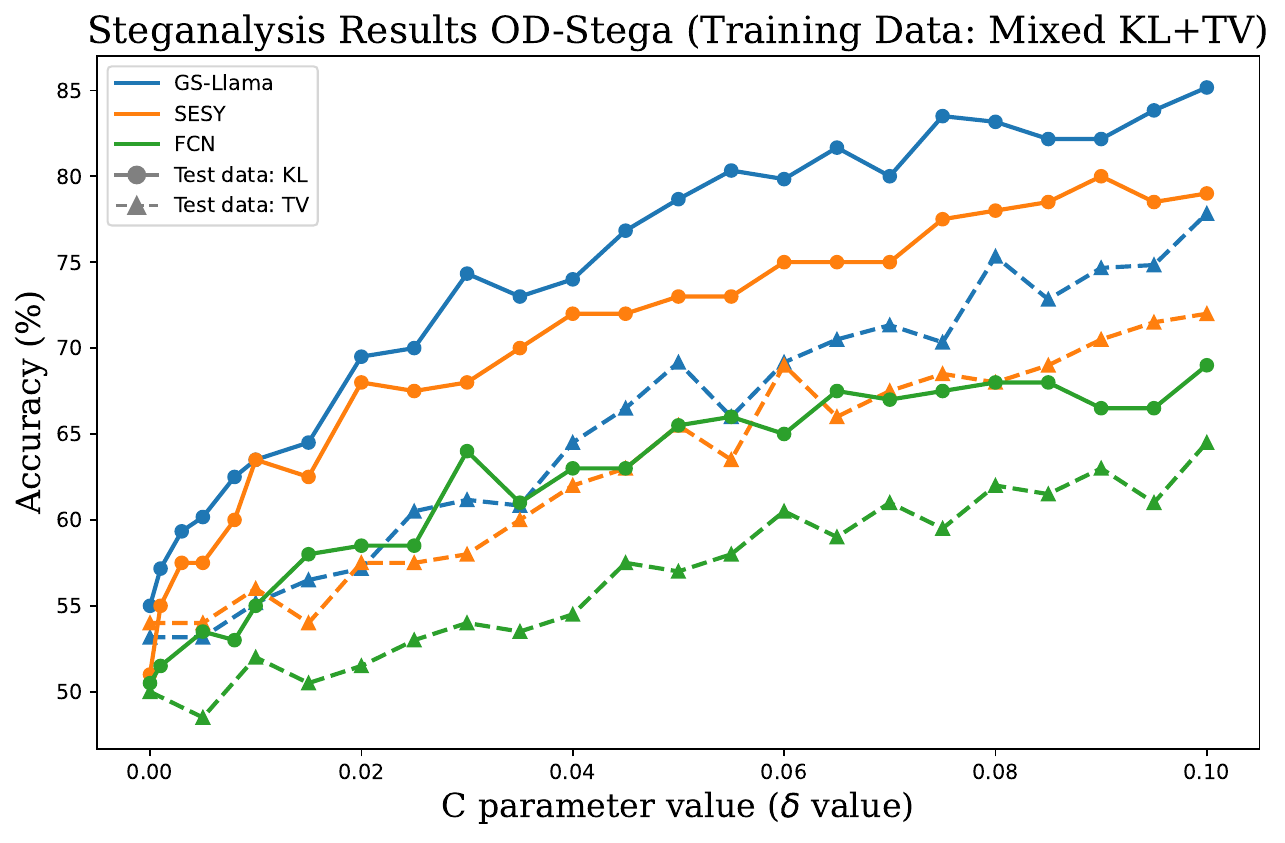}
    \caption{Steganalysis results of three detection models on OD-Stega, with accuracy plotted on the y-axis.}
    \label{fig:od-detection-acc}
\end{figure}

\begin{figure}[ht]
    \centering
    \includegraphics[width=\linewidth]{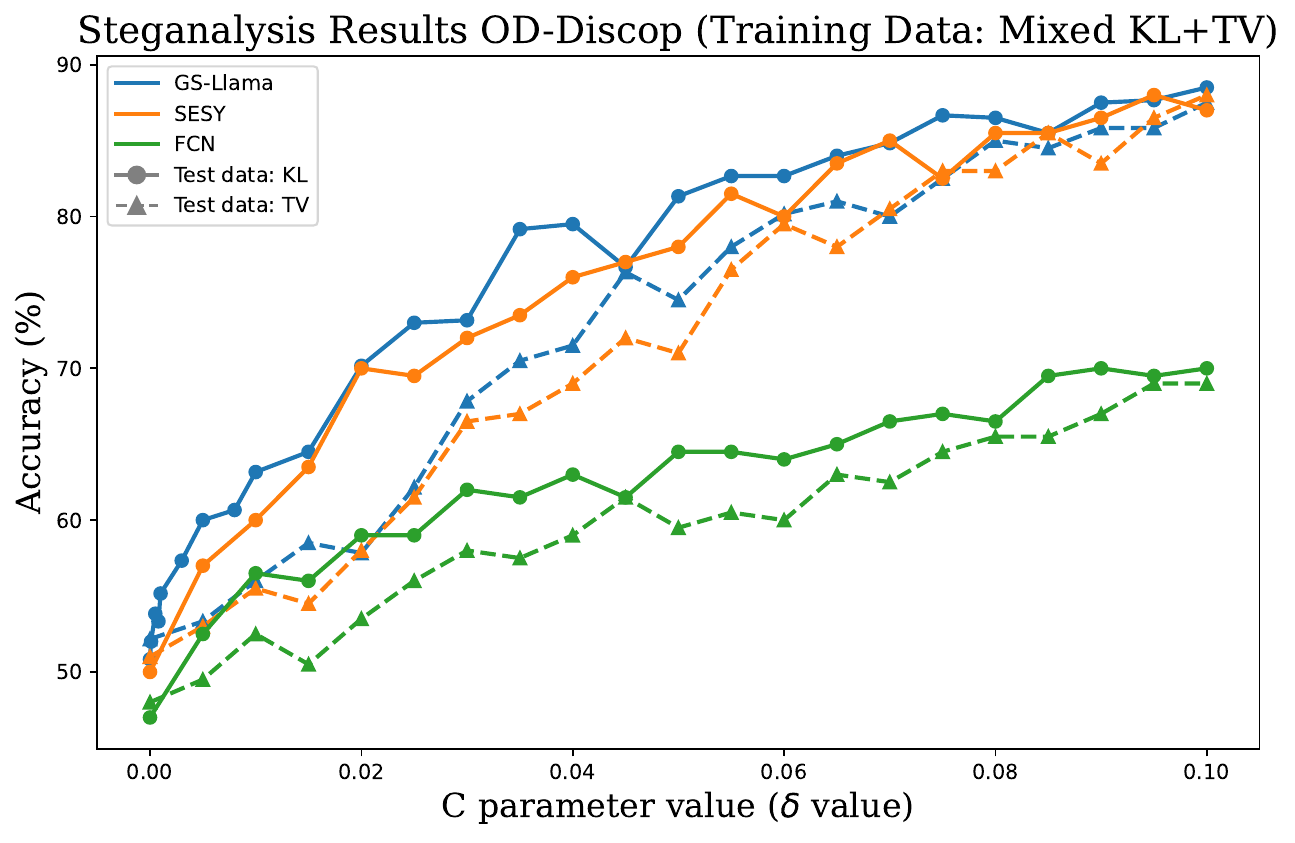}
    \caption{Steganalysis results of three detection models on OD-Discop, with accuracy plotted on the y-axis.}
    \label{fig:discop-detection-acc}
\end{figure}

In this section, we present the accuracy plots of steganalysis results for the three models: FCN, SESY, and GS-LLaMA. As shown in Figure \ref{fig:od-detection-acc} and Figure \ref{fig:discop-detection-acc}, it is evident that FCN achieves lower accuracy compared to the other two models. In fact, FCN struggles to reach 70\% accuracy, even when the parameter \( C \) is set to a relatively large value such as 0.1.
In contrast, the other two models—SESY and GS-LLaMA achieve over 70\% accuracy when \( C > 0.05 \). This improved performance is likely due to their transformer-based architectures. In particular, GS-LLaMA, which is fine-tuned from the LLaMA3-8B model, serves as a very strong detector in this context.

\section{Parameter Choice to Avoid Tokenization Error and Experiment Details}
\label{sec:token}
We heuristically determine the length of $B$ that guarantees the steganography process succeeds with high probability, which we set as $1-10^{-8}$ in our work. We observe that for Llama2-7B models, a single bit produces a tokenization error at a rate below $2 \times 10^{-4}$. We can make $2^{|B|}$ separate attempts, and only one successful attempt is needed, i.e., 
\begin{align}
& 1 - \left( (|B|+|S|) \times 2 \times 10^{-4} \right) ^{2^{|B|}} \geq 1-10^{-8} \\
\Rightarrow & |B| > 3 - \log_2 \left( 4 -\log_{10}2 - \log_{10}|S|\right) \label{eq:ignore_b_in_b+L},
\end{align}
assuming $B$ is significantly shorter than $S$. Bob can indeed determine the length of $B$ by referencing the length of the secret message $S$ using equation (\ref{eq:ignore_b_in_b+L}). 

Our experiment begins with random B-bit prefixes, which is equivalent to selecting a random initial interval in the probability distribution, ensuring that token sequence selection is unbiased under this process. If this generated text runs into a tokenization error, we randomly sample a new B-bit prefix. In our experiments, out of 100 generated text sequences of 25 tokens each, fewer than 10 cases needed a second attempt with a new prefix B, showing that the iteration rate of this procedure is sufficiently low.

\section{Non-Stego vs. $\delta = 0$ Stego-Text} \label{sec:NSvsdelta0}

We employ GS-Llama as a classifier to assess whether the detector can differentiate between natural text and stego-text encoded with $\delta = 0$. For this purpose, we train a binary classification approach using a balanced dataset of 1,000 samples, consisting of 500 naturally generated texts (from Llama2-7B) and 500 $\delta = 0$ OD-stage encoded texts. The stego-texts were labeled as "stego," while natural texts were labeled as "non-stego". Each input followed a standardized prompt format:

\begin{verbatim}
### Text:
{input_text}

### Question:
Is the above text stego or
non-stego?

### Answer:
\end{verbatim}

We fine-tune the Llama-3.1-8B-Instruct model to function as a classifier with the same training details provided in section \ref{sec:detection_setup} and assess its performance using a separate test set. As shown in Table \ref{tab:NSvsdelta0_result1} and \ref{tab:NSvsdelta0_result2}, the classifier achieve an overall accuracy of 49.8\% and a macro F1 score of 49.8\%, performing at chance level. These results indicate that $\delta = 0$ stego-texts are computationally indistinguishable from naturally generated texts, even when analyzed by a strong large language model-based detector.

\begin{table}[h]
\caption{Binary classifier results on 500 stego-texts and 500 natural texts. The value in the table is the number of texts count.}
\centering
\begin{tabular}{l|l|l}
\diagbox{True}{Predict} & Stego & Non-stego \\ \hline
\begin{tabular}[c]{@{}l@{}}Stego $\delta = 0$\end{tabular} & 248   & 252       \\ \hline
Non-stego                                                    & 250   & 250      
\end{tabular}
\label{tab:NSvsdelta0_result1}
\end{table}

\begin{table}[h]
\caption{Binary classifier results on 500 stego-texts and 500 natural texts. These are the common metrics that evaluate the model performances. }
\centering
\begin{tabular}{l|l|l|l}
Test data & Precision & Recall & F1 \\ \hline
\begin{tabular}[c]{@{}l@{}}Stego $\delta = 0$\end{tabular} & 0.498  & 0.496 & 0.497       \\ \hline
Non-stego                                                    & 0.498   & 0.50 & 0.499     
\end{tabular}
\label{tab:NSvsdelta0_result2}
\end{table}

\section{Additional GPT Evaluation Result} \label{sec:GPT_eval_results}

\begin{figure}[h]
    \includegraphics[width=\columnwidth]{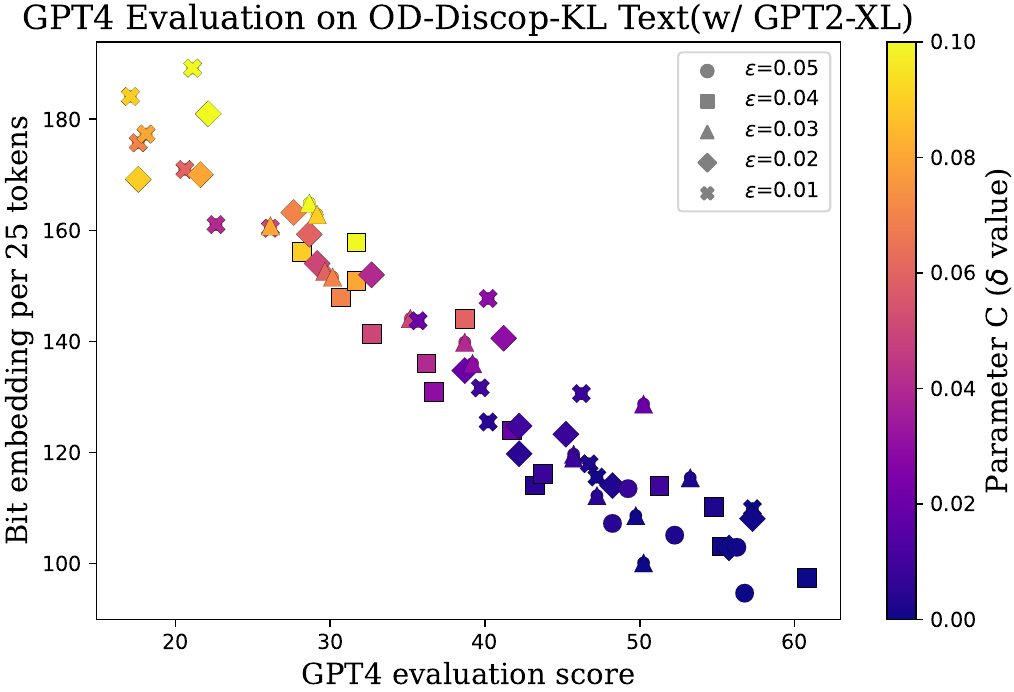}
    \caption{GPT4 detection score on OD-Discop-KL with GPT2-XL vs. average (over 200 texts) bits embedded per 25 tokens.}
    \vspace{-0.3cm}
    \label{fig:GPT4-eval-discop-kl}
\end{figure}
\begin{figure}[h]
    \includegraphics[width=\columnwidth]{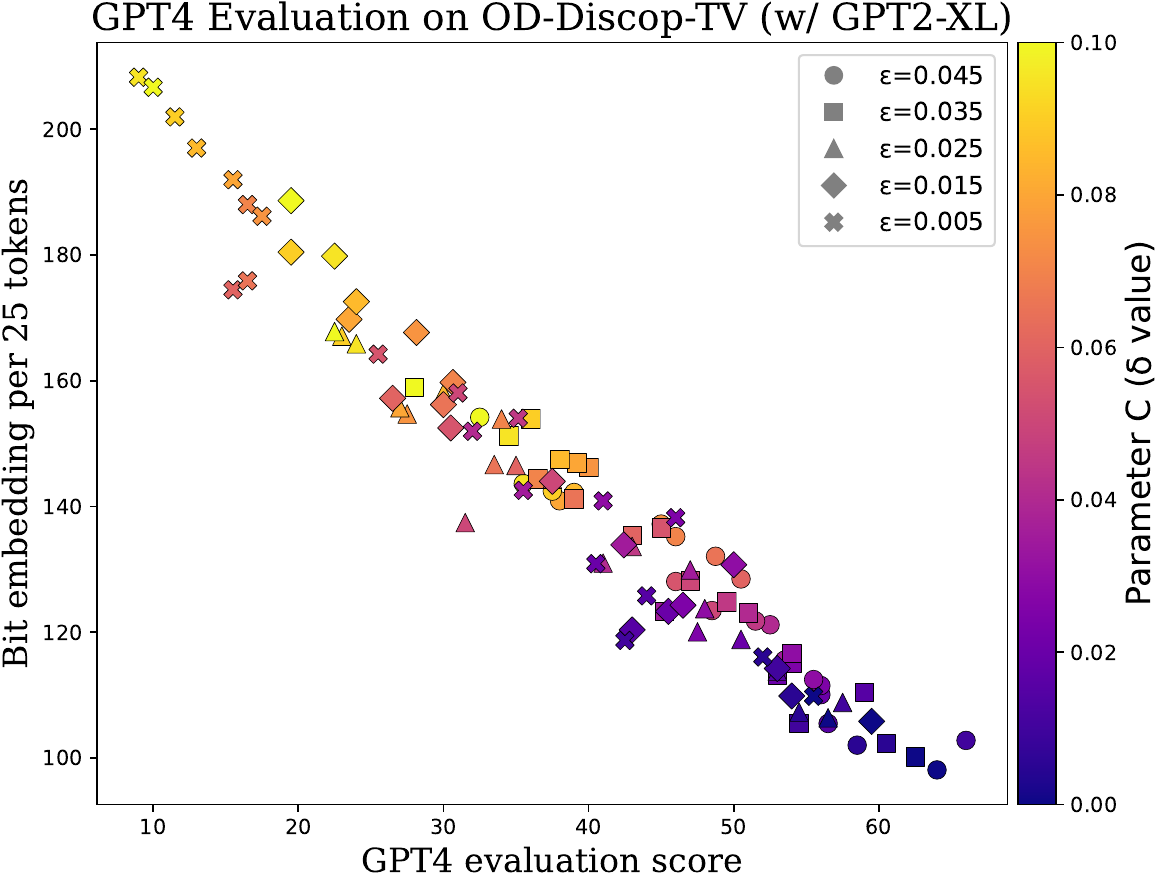}
    \caption{GPT4 detection score on OD-Discop-TV with GPT2-XL vs. average (over 200 texts) bits embedded per 25 tokens.}
    \vspace{-0.3cm}
    \label{fig:GPT4-eval-discop-tv}
\end{figure}

We include the result of GPT-4 evaluation on OD-Discop in Figure \ref{fig:GPT4-eval-discop-kl} and \ref{fig:GPT4-eval-discop-tv}.

\section{Steganalysis Result on Longer Text Sequence}\label{sec:detection_tk100}
\begin{figure}[ht]
    \centering
    \includegraphics[width=\linewidth]{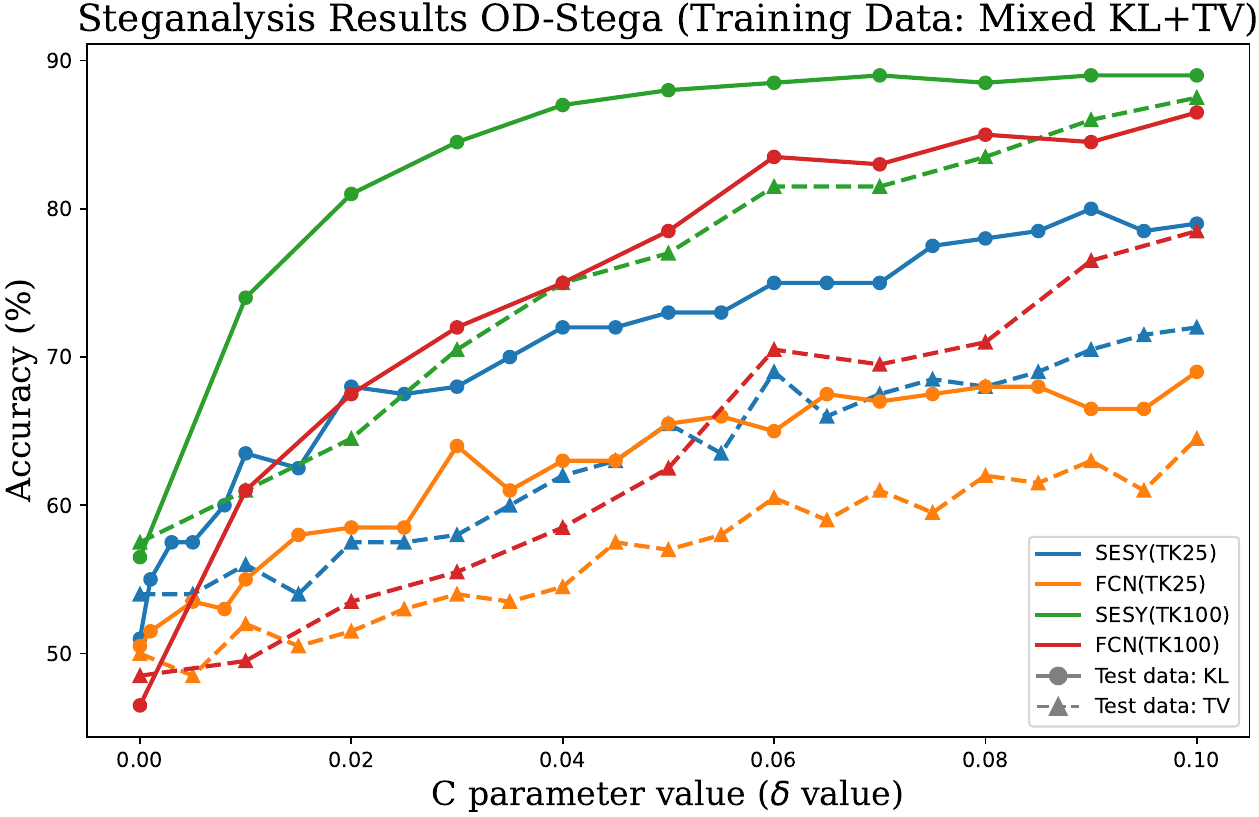}
    \caption{Steganalysis results on longer text sequences (token length 100) verses shorter text sequences (token length 25), with accuracy plotted on the y-axis.}
    \label{fig:OD-tk100-steganalysis-result}
\end{figure}

We retrained the detection models with extended text sequences and conduct steganalysis on the OD-stega outputs with token length of 100, using a probability cutoff of $\epsilon = 0.025$ and varying the parameter C in the range from $0.01$ to $0.1$. The results are presented in Figure \ref{fig:OD-tk100-steganalysis-result}. In both detection methods, the accuracy curves follow the same trend as those for the shorter sequences but are shifted upward. This upward shift occurs because longer sequences make the embedded signal easier to detect: as the detector observes more tokens, subtle distributional differences in the stego-text accumulate and become more perceptible. These results further motivate embedding more information into shorter texts to reduce detectability.

\begin{figure}[ht]
    \centering
    \includegraphics[width=\linewidth]{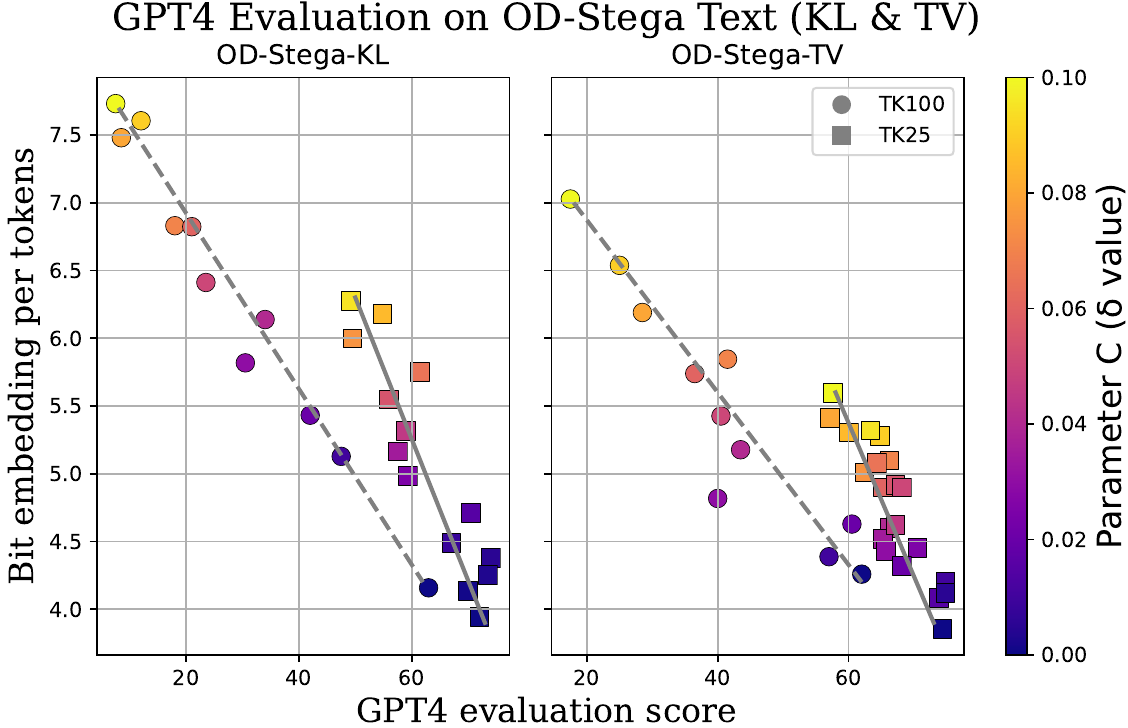}
    \caption{GPT4 detection score on OD-Stega longer text sequences (token length 100) verses shorter text sequences (token length 25), with average embedding rate on the y-axis.}
    \label{fig:OD-tk100-GPT4-results}
\end{figure}

Figure \ref{fig:OD-tk100-GPT4-results} presents a comparison of GPT-4 evaluation scores for token length 100 (circular markers) and token length 25 (square markers). The dashed line and solid line illustrate how the scores for token 100 and token 25, respectively, evolve as the parameter $C$ increases. We observe that as $C$ increases, the distance between the two lines becomes larger, indicating that the divergence of each token accumulates over the length of the texts. Consequently, text sequences of length 100 exhibit a higher average embedding rate but a lower detection score, implying they are more likely to be flagged as abnormal. This observation is consistent with the results shown in Figure \ref{fig:OD-tk100-steganalysis-result}.

\section{More Graphs}
\begin{figure}[t]
    \centering
    \includegraphics[width=\linewidth]{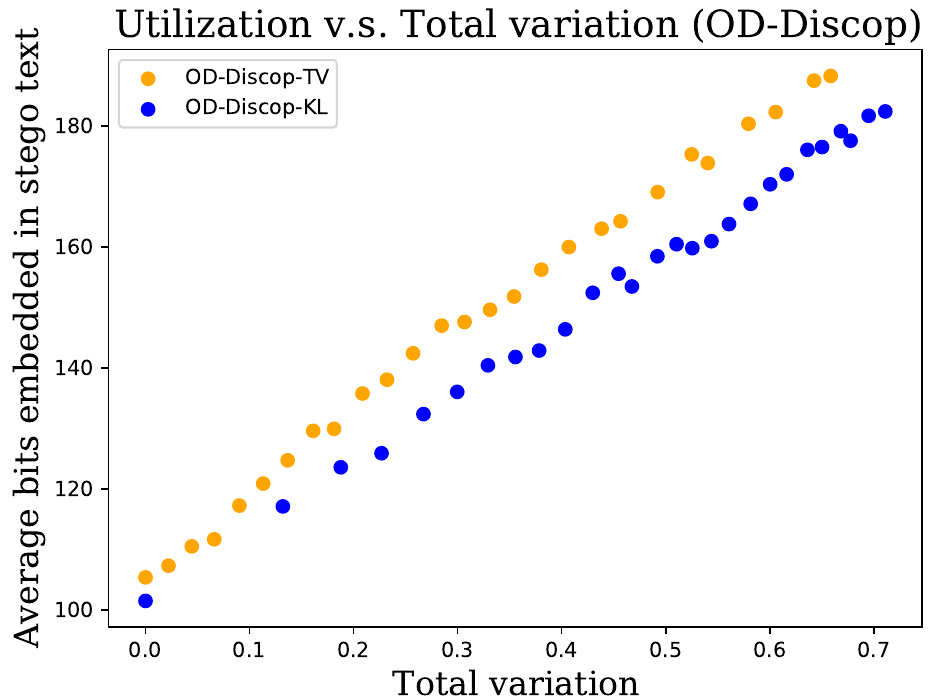}
    \caption{Average bits embedded using different adjusting methods OD-Discop-KL and OD-Discop-TV under the same total variation distance.}
    \label{fig:bits_vs_tv_discop_kl+tv}
\end{figure}

In Figure \ref{fig:bits_vs_tv_discop_kl+tv}, we present the embedding utilization behavior under a fixed total variation constraint for OD-Discop stego-texts. While OD-Discop-KL (or temperature adjustment) is optimal for maximizing the number of embedded bits under a KL divergence constraint, it is not optimal under a TV constraint. This indicates that temperature tuning is no longer the best adjustment strategy when a different constraint, such as total variation, is considered. 

\section{More Stego-Text Examples} \label{sec:moreexamples}
Tables \ref{tab:example_text2} and \ref{tab:example_text3} demonstrate the generated OD-KL stego-text by setting the parameter \(\epsilon = 0.025\) and comparing results from small and larger adjustment values $C$. In Table \ref{tab:example_text3}, odd content arises in the larger parameter \(C = 0.05\) stego-text. Nonetheless, in most instances, the stego-text generated with a larger \(C\) appears typical, as depicted in Table \ref{tab:example_text2}.

\begin{table}[h]
\centering
\caption{Example stego-texts OD-KL: Same cutoff, different adjustment parameter $C$.}
\vspace{-0.3cm}
\renewcommand{\arraystretch}{1.2}
\scriptsize
\begin{tabularx}{\columnwidth}{l|X|l|l}
\toprule
Parameters                                                           & Prompt: Once upon a time, there was a princess who & SESY & FCN \\ 
\midrule
\begin{tabular}[c]{@{}l@{}}$C = 0.01$\\ $\epsilon = 0.025$\end{tabular} &was going to marry a prince. She lived in a cramped palace. It wasn’t big, and she wasn… & NS & NS \\ \hline
\begin{tabular}[c]{@{}l@{}}$C = 0.05$\\ $\epsilon = 0.025$\end{tabular} & liked travel. After a few sprees through Paris, Milan, Rome and Palm Springs, she… & S& NS \\ \hline
\begin{tabular}[c]{@{}l@{}}$C = 0.09$\\ $\epsilon = 0.025$\end{tabular} & cried  out, 'Ah well, I do adore  thick pungent sweat pouring down onto my crim… & S& S \\ 
\bottomrule
\end{tabularx}
\label{tab:example_text2}
\end{table}

\begin{table}[h]
\centering
\caption{Example stego-texts OD-KL: Same cutoff, different adjustment parameter $C$.}
\vspace{-0.3cm}
\renewcommand{\arraystretch}{1.2}
\scriptsize
\begin{tabularx}{\columnwidth}{l|X|l|l}
\toprule
Parameters                                                           & Prompt: Over the next few days, the weather will be   &  SESY & FCN \\ 
\midrule
\begin{tabular}[c]{@{}l@{}}$C = 0.01$\\ $\epsilon = 0.025$\end{tabular} &  16 to 20 degrees, after the 11th~12th temperature was slightly lower,… & NS & NS  \\ \hline
\begin{tabular}[c]{@{}l@{}}$C = 0.05$\\ $\epsilon = 0.025$\end{tabular} & 4 minutes earlier each night, Sun/Mon AFAIK no Co2 Med Decision Don, MS October 9… & S & S  \\
\bottomrule
\end{tabularx}
\label{tab:example_text3}
\end{table}

In Table \ref{tab:example_text_ODTV}, we present examples produced by OD-TV adjustments. Notably, SESY demonstrates its strength. Even when the text appears normal in the $C=0.05$ scenario, SESY still identifies it as stego-text correctly.

In tables \ref{tab:OD-Discop_exp1} and \ref{tab:OD-Discop_eg2}, we have included several text samples that illustrate the OD-Discop method. In these examples, the cutoff value $\epsilon$ is held constant while the value of the parameter $C$ is varied. It is evident that in certain prompts the OD-Discop texts continue to appear normal even when the parameter $C$ is considerably high. It is observed that this occurs because such prompts are capable of generating text sequences with higher entropy. Tables also show that SESY outperforms FCN as a detector.

\begin{table}[h]
\centering
\caption{Example stego-texts OD-TV: Same cutoff, different adjustment parameter C}
\vspace{-0.3cm}
\renewcommand{\arraystretch}{1.2}
\scriptsize
\begin{tabularx}{\columnwidth}{l|X|l|l}
\toprule
Parameters                                                           & Prompt:  I went to this restraunt the other day, and I would rate its food  &  SESY & FCN \\ 
\midrule
\begin{tabular}[c]{@{}l@{}}$C = 0.01$ \\ $\epsilon = 0.025$\end{tabular} & 4 stars out of 5. The place was half empty, which was great, because the only thing you needed to fill & NS & NS \\ \hline
\begin{tabular}[c]{@{}l@{}}$C = 0.03$\\ $\epsilon = 0.0175$\end{tabular} &  8/10. My mother and father said however, that its not as good as it used to be. Never ce... & NS & NS \\ \hline
\begin{tabular}[c]{@{}l@{}}$C = 0.05$\\ $\epsilon = 0.005$\end{tabular} & 6 out of 10. It was dirty and dishevelled, too. The load text and video entered... & S & NS \\ 
\bottomrule
\end{tabularx}
\label{tab:example_text_ODTV}
\end{table}

\begin{table}[h]
\centering
\caption{Example stego-texts: OD-Discop-KL stego-text example (GPT2-XL).}
\vspace{-0.3cm}
\renewcommand{\arraystretch}{1.2}
\scriptsize
\begin{tabularx}{\columnwidth}{l|X|l|l}
\toprule
Parameters                                                           & Prompt:  There are many species of animals living in the Amazon rainforest, including species such as  &SESY&FCN  \\ 
\midrule
\begin{tabular}[c]{@{}l@{}} Discop \\ $\epsilon = 0.025 $\end{tabular} & ursids such as Amazon fire bats, but the recently announced discoveries in China of evidence for alien primates was an "earth-... & NS & NS \\ \hline
\begin{tabular}[c]{@{}l@{}}$C = 0.01$\\ $\epsilon = 0.025$\end{tabular} & ichthyosaurs, which were partly scavenged by large turtles like ammonoids, which had a diet that consisted mostly of wood... & NS & NS \\ \hline
\begin{tabular}[c]{@{}l@{}}$C = 0.05$\\ $\epsilon = 0.025$\end{tabular} & ichthyosaurs that sat 150 feet underground, sometimes buried in silt between a surface rainforest (corals and mounds... & S &NS \\ \hline
\begin{tabular}[c]{@{}l@{}}$C = 0.09$\\ $\epsilon = 0.025$\end{tabular} & ichneumon beetles, carbon dioxide-loving methanodonts, rodents, humans and landrover loads of other... & S & NS \\ \hline
\toprule
Parameters                                                           & Prompt:  For all the sports fans out there, there was a recent upset between  &  \\ \midrule 
\begin{tabular}[c]{@{}l@{}} Discop \\ $\epsilon = 0.025 $\end{tabular} & Panthers LB Luke Kuechly and Saints LB Joe Vellano, in the same division. If Carolina... & NS & NS \\ \hline
\begin{tabular}[c]{@{}l@{}} $C = 0.01$ \\ $\epsilon = 0.025 $\end{tabular} & iced tea and hot water that you really probably missed. Speaking of hot, that Minnesota Snack Club (someone hurt someone)... & NS & S \\ \hline
\begin{tabular}[c]{@{}l@{}} $C = 0.03$ \\ $\epsilon = 0.025 $\end{tabular} & Kent State and Utah.   What I really miss most about the neutral court, and my capacity to sit and watch sports... & NS &NS \\ \hline
\begin{tabular}[c]{@{}l@{}} $C = 0.07$ \\ $\epsilon = 0.025 $\end{tabular} & Mount Bank BC and Carrie 10 BC that saw the band and its supporters are disillusioned with Season 5!!! Let me re ... & S &S \\ \hline
\begin{tabular}[c]{@{}l@{}} $C = 0.09$ \\ $\epsilon = 0.025 $\end{tabular} & lloVici Gaming and Nomia PK brand. The flag-ship Jungler hit kills around the map in the bad Fight... & S &NS \\ \hline
\end{tabularx}
\label{tab:OD-Discop_exp1}
\end{table}

\begin{table}[t]
\centering
\caption{Example stego-texts: OD-Discop-TV stego-text example (GPT2-XL).}
\vspace{-0.3cm}
\renewcommand{\arraystretch}{1.2}
\scriptsize
\begin{tabularx}{\columnwidth}{l|X|l|l}
\toprule
Parameters                                                           & Prompt:  In this blog post, I would like to recount an event that happened to me the other day. I was leaving my house when  &  SESY & FCN \\ 
\midrule
\begin{tabular}[c]{@{}l@{}} Discop \\ $\epsilon = 0.025 $\end{tabular} & I came across an interesting volume of mythography. The celestial mythographers of Mesopotamia were present in very large... & S & S \\ \hline
\begin{tabular}[c]{@{}l@{}}$C = 0.01$\\ $\epsilon = 0.025$\end{tabular} &  I noticed things very strangely going on inside my front door. Something was big and metallic, so it seemed to be drawing... & NS & NS \\ \hline
\begin{tabular}[c]{@{}l@{}}$C = 0.03$\\ $\epsilon = 0.025$\end{tabular} & as I rounded the corner to the tenway exit I saw a taxi cab driver swatting at a debt firm car out... & S & NS \\ \hline
\begin{tabular}[c]{@{}l@{}}$C = 0.05$\\ $\epsilon = 0.025$\end{tabular} & I felt hering Chinese launched his strike at me. It all happened very quickly and in that moment I just realised I... & S & NS \\ \hline
\begin{tabular}[c]{@{}l@{}}$C = 0.07$\\ $\epsilon = 0.025$\end{tabular} & a pigeon appeared from outside, RITP puffing till it enlarged the bolt holes on my roof! A third... & S & NS\\ \hline
\toprule
Parameters                                                           & Prompt:  In the recent Tokyo 2024 Olympics, the most notable event was the  &  \\ 
\midrule
\begin{tabular}[c]{@{}l@{}} Discop \\ $\epsilon = 0.025 $\end{tabular} & ichinen sumo wrestling contest between sumo's three greatest battlers, Mark Kerr, Babaev and Hikaru S... & NS &NS\\ \hline
\begin{tabular}[c]{@{}l@{}}$C = 0.01$\\ $\epsilon = 0.025$\end{tabular} &  ichi rugby sevens qualifying match between Britain, Scotland and Fiji, whose fans underlined their policy of joining in whenever Japan has... & NS & NS \\ \hline
\begin{tabular}[c]{@{}l@{}}$C = 0.03$\\ $\epsilon = 0.025$\end{tabular} & izakaya ("bar and sauna") being erected in Ho-ho-koo, right next to Mount Fuji,... & NS & NS  \\ \hline
\begin{tabular}[c]{@{}l@{}}$C = 0.05$\\ $\epsilon = 0.025 $\end{tabular} & izakaya span will open to bid patrons starting 15 months after the November 2015 opening day, when you would likely fail the... & S & NS \\ \hline
\end{tabularx}
\label{tab:OD-Discop_eg2}
\end{table}

\begin{figure*}[h]
    \centering
    \includegraphics[width=1\textwidth]{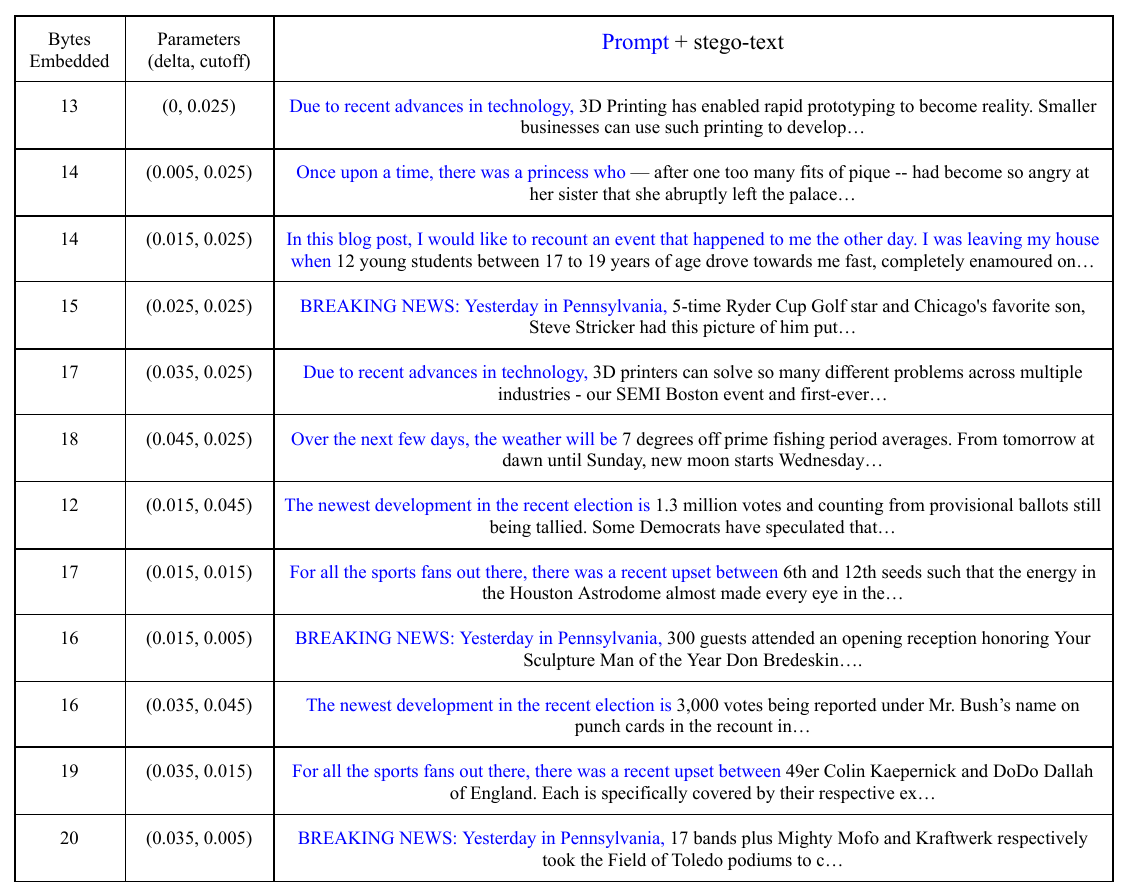}
    \caption{Stego-text examples (OD-KL) in different pair of parameters $(C, \epsilon)$ and length of secret message embedded.}
    \label{fig:longtext2}
\end{figure*}

\begin{figure*}[h]
    \centering
    \includegraphics[width=1\textwidth]{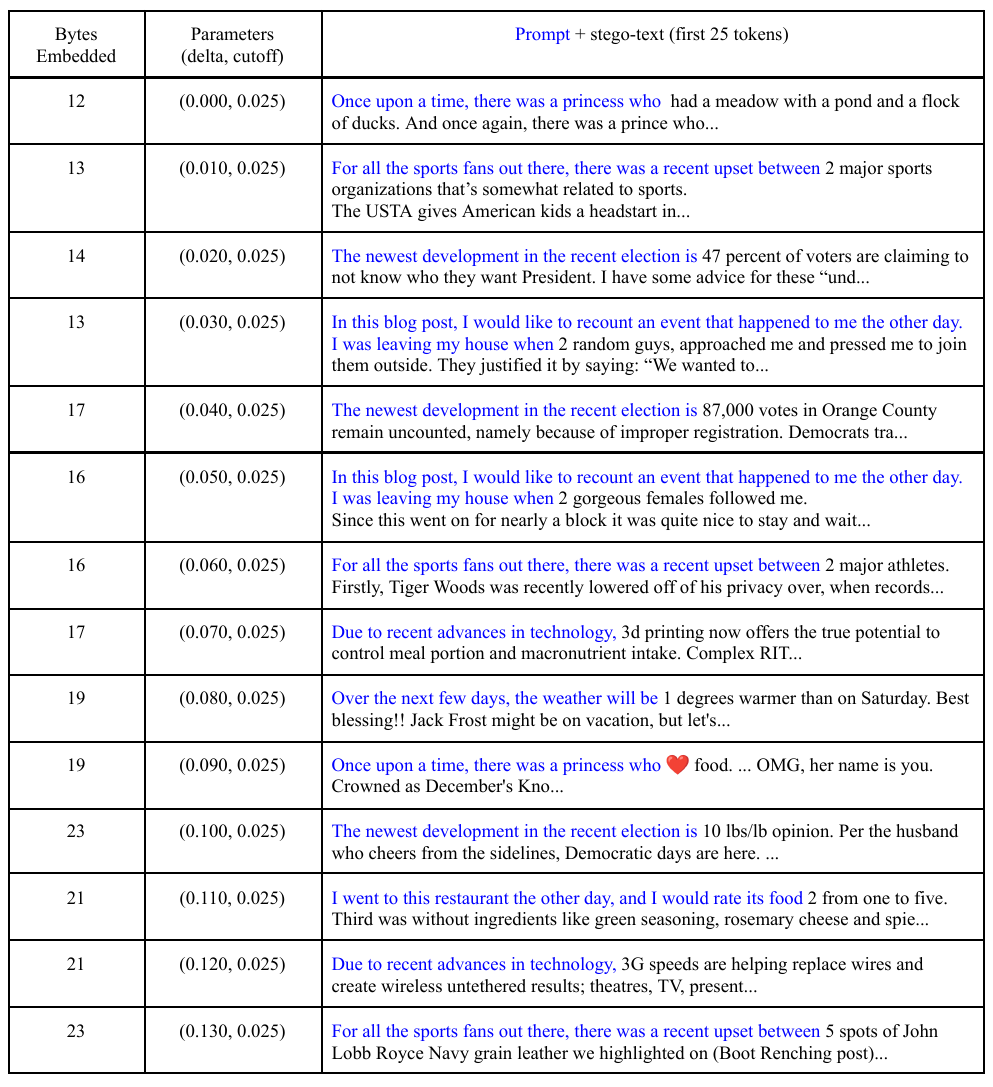}
    \caption{Stego-text examples (OD-TV) in different pair of parameters $(C, \epsilon = 0.025)$ and length of secret message embedded.}
    \label{fig:longtext3}
\end{figure*}

In Figure \ref{fig:longtext2} and \ref{fig:longtext3} we show more examples of generated stego-texts using the proposed OD-Stega approach with various parameters. It can be seen that as the $C$ parameters increase for fixed cutoff value, the embedding capability increases. The generated stego-texts mostly remain fluent in this parameter range.

\newpage

\end{appendices}
\end{document}